\documentclass[12pt]{article}
\usepackage{latexsym}
\usepackage{epsfig,amssymb,euscript}
\usepackage{amsmath}
\usepackage{array,calc,epsfig}
\usepackage{citesort}

\def\be{\begin{equation}}
\def\ee{\end{equation}}
\def\bea{\begin{eqnarray}}
\def\eea{\end{eqnarray}}

\numberwithin{equation}{section} 
\usepackage[]{graphicx}

\def\cala         {{\cal A}}

\def\calb         {{\cal B}}
\def\calc         {{\cal C}}

\def\cale         {{\cal E}}
\def\calf         {{\cal F}}
\def\calg         {{\cal G}}

\def\calm         {{\cal M}}
\def\caln         {{\cal N}}

\def\d              {\text{d}}

\def\fraks              {{\mathfrak S}}
\def\frakc               {{\mathfrak C}}
\def\frakf               {{\mathfrak F}}

\def\p{{}^{\prime}}

\def\Z          {{\mathbb Z}}

\def\hsp{,\hspace{.7cm}}

\def\Re           {{\rm Re\hskip0.1em}}
\def\Im           {{\rm Im\hskip0.1em}}

\def\H{ \hbox{\rm H}}

\def\Im{ \hbox{\rm Im}}
\def\Ker{ \hbox{\rm Ker}}

\def\sqr#1#2{{\vcenter{\vbox{\hrule height.#2pt
 \hbox{\vrule width.#2pt height#1pt \kern#1pt \vrule width.#2pt}\hrule
 height.#2pt}}}}

\def\Generalized{Generalized}
\def\generalized{generalized}

\def\slashchar#1{\setbox0=\hbox{$#1$}           
\dimen0=\wd0                                 
\setbox1=\hbox{/} \dimen1=\wd1               
\ifdim\dimen0>\dimen1                        
\rlap{\hbox to \dimen0{\hfil/\hfil}}      
#1                                        
\else                                        
\rlap{\hbox to \dimen1{\hfil$#1$\hfil}}   
/                                         
\fi}

\begin{document}
\font\cmss=cmss10 \font\cmsss=cmss10 at 7pt
\leftline{\tt hep-th/0703129}

\vskip -0.5cm
\rightline{\small{\tt KUL-TF-07/06}}
\rightline{\small{\tt ULB-TH/07-12}}

\def\thefootnote{\fnsymbol{footnote}}

\vskip .7 cm

\hfill
\vspace{13pt}
\begin{center}
{\Large \textbf{D-brane networks in flux vacua,\\ generalized cycles and calibrations}}
\end{center}

\vspace{5pt}
\begin{center}
{\large\textsl{Jarah Evslin$^a$\footnote{\texttt{ jevslin@ulb.ac.be}} and Luca Martucci~$^{b}$\footnote{\texttt{ luca.martucci@fys.kuleuven.be}}}}
\end{center}
\renewcommand{\thefootnote}{\arabic{footnote}}

\begin{center}
\vspace{1em}
{\em  {\small $^a$ International Solvay Institutes,\\
Physique Th\'eorique et Math\'ematique,\\
Statistical and Plasma Physics C.P. 231,\\
Universit\'e Libre
de Bruxelles, \\ B-1050, Bruxelles, Belgium\\}}

\vspace{18pt}
\textit{\small $^b$ Institute for Theoretical Physics, K.U. Leuven,\\ Celestijnenlaan 200D, B-3001 Leuven, Belgium}\\  \vspace{6pt}
\end{center}

\vspace{11pt}

\begin{center}
\textbf{Abstract}
\end{center}

\noindent

We consider chains of generalized submanifolds, as defined by Gualtieri in the context of generalized complex geometry, and define a boundary operator that acts on them. This allows us to define generalized cycles and the corresponding homology theory. Gauge invariance demands that D-brane networks on flux vacua must wrap these generalized cycles, while deformations of generalized cycles inside of a certain homology class describe physical processes such as the dissolution of D-branes in higher-dimensional D-branes and MMS-like instantonic transitions. We introduce calibrations that identify the supersymmetric D-brane networks, which minimize their energy inside of the corresponding homology class of generalized cycles. Such a calibration is explicitly presented for type II $\caln=1$ flux compactifications to four dimensions.  In particular networks of walls and strings in compactifications on warped Calabi-Yau's are treated, with explicit examples on a toroidal orientifold vacuum and on the Klebanov-Strassler geometry.

%




\newpage
\setcounter{footnote}{0}


\tableofcontents




\vspace{1.5cm}

\section{Introduction}

Consider a flux vacuum of type II string theory.  A nontrivial Neveu-Schwarz (NS) $H$ flux can give rise to strong constraints on the allowed D-brane configurations \cite{FW}, since the pullback of $H$ to a D-brane worldvolume must be exact.  At the same time, a D-brane cannot wrap a submanifold with boundary, since this would violate Ramond-Ramond (RR) gauge invariance. Both of these pathologies can be cured by letting end, let us say,  D$(p-2)$-branes on a D$p$-brane, D$(p-4)$-branes on a D$(p-2)$ and so on \cite{strominger,townsend,mms}.  Thus, from the topological point of view, in the presence of background and worldvolume fluxes it is natural to consider networks of D-branes of different dimensions. 

On the other hand, from the dynamical point of view, a satisfactory description of such configurations is obviously problematic without further assumptions. Experience teaches us that supersymmetry can simplify life considerably, while still leaving room for nontrivial and physically relevant effects. In the absence of fluxes, a single static brane wraps a volume minimizing cycle and indeed supersymmetric D-branes correspond to the classical notion of calibrated cycles \cite{calor}, where the background calibration (a closed form) is directly related to the background supersymmetry (see for example \cite{gauntlett}). The inclusion of background fluxes has led \cite{papa} to the generalization of the classical notion of a calibration  from a volume minimizing to an energy minimizing background form. Following the same philosophy, it was shown in \cite{paulk,luca1} that the definition of calibration for D-branes in type II theories can be further extended to automatically include a nontrivial background $H$-flux and the worldvolume field strength $\calf$.  In \cite{paulk,luca1} these generalized calibrations were found not to be differential forms of definite degree, but rather {\em polyforms}, i.e. sums of forms of different degrees, which characterize calibrated  supersymmetric D-branes of different dimensions. 

The setting of \cite{paulk,luca1} immediately suggests its applicability to D-brane networks instead of single D-branes, and this indeed would constitute its more natural formulation in light of the topological arguments discussed above. In this paper we will develop such an approach and see that it offers a new unifying formalism that efficiently describes many kinds of D-brane networks. 
As an ordinary calibration identifies the volume minimizing cycle inside of a certain homology class, to discuss calibrated D-brane networks we must define what is the proper homological equivalence that relates D-brane networks that are in some sense continuously connected. First of all, it is natural to use the generalized submanifolds of generalized complex geometry (GCG) \cite{hitchin,gualtieri} where one considers a pair $(\Sigma,\calf)$ given by a submanifold $\Sigma$ and a worldvolume field strength $\calf$ defined on it. Indeed the relation between D-brane physics on flux vacua and generalized (complex) geometry, already suggested by the use of polyforms, has deep implications \cite{paulk,luca1,luca2,gmeiner,lucadef}. Second, generalized chains are weighted sums of generalized submanifolds, possibly of different dimensions. We will show how a proper generalized boundary operator acting on the generalized chains can be defined. As we will discuss in detail, the resulting homology theory will describe the different kinds of possible transitions between D-brane configurations: ordinary homological deformations, dissolutions of D-branes in higher dimensional branes \cite{Myers},  decay or creation  of D-branes via the nucleation of higher dimensional Euclidean branes \cite{mms}, and so on.   Thus, the generalized calibration defines the stable, energy minimizing, D-brane networks taking into account all of these kinds of possible transitions.  

The generalized boundary operator defines a homology group that naturally classifies different D-brane networks. The problem of the topological classification of D-brane networks will not be the central one in this paper. However it is an important one and for this reason we relegate a discussion on it to appendix A, where we relate the homology discussed above to the more standard twisted homology discussed e.g. in \cite{Andres}.   

The main example in which we apply our general approach is the class of type II $\caln=1$ D-calibrated flux vacua considered in \cite{luca1,luca2,lucadef} (for a short review see \cite{lucarev}).  These backgrounds preserve four-dimensional Poincar\'e invariance and admit supersymmetric D-branes which are spacetime filling, domain walls or strings in the four flat dimensions, or networks of them. We construct the appropriate generalized calibration that allows us to treat all these different configurations at once, and we apply it to the case of IIB flux compactifications on warped Calabi-Yau manifolds.

We begin in Sec.~\ref{sec1} by defining generalized chains and their duals, generalized currents.  We define a generalized boundary operator $\hat\partial$ on these chains which we use to define a homology theory.  We see that gauge-invariance requires that D-brane networks wrap generalized cycles.  Then in Sec.~\ref{genhom} we interpret generalized boundaries as generalized cycles wrapped by unstable networks.  This implies that networks which represent the same $\hat\partial$-homology class may be related by dynamical processes.  We give several examples, such as branes dissolving in branes and the MMS-like  \cite{mms} instantonic processes like that of Kachru, Pearson and Verlinde \cite{kpv}.  In Sec.~\ref{calisec} we move beyond topology, and try to understand some of the dynamics of networks.  We define a calibration that is well-defined on all generalized cycles such that networks minimize their energy if and only if they wrap calibrated generalized cycles.  As a simple example, we apply this technology to reproduce the BIon solution \cite{BIon}.  In Sec.~\ref{fluxsec} we specialize the previous discussion to $\mathcal{N}=1$ flux compactifications, in which we find a single generalized calibration that calibrates string, domain wall and space-filling D-branes in the 4-dimensional theory.  This discussion is further specialized in Sec.~\ref{cysec} to the case of compactifications on type IIB warped Calabi-Yau's, where we discuss some general properties of supersymmetric networks of domain walls at different angles glued together along a string-like common boundary.  Explicit examples are given of junctions of walls in the toroidal orientifold compactification $T^6/\mathbb{Z}_2$ flux vacua introduced in \cite{kachru}.  In Sec.~\ref{2exsec} we explicitly solve the BPS equations describing the geometry of composite domain walls, where a number of space-time filling D3-branes end on a D5-brane wrapping  an internal three-cycle supporting a non-trivial $H$-flux.   Two concrete examples are considered, one in the aforementioned toroidal orientifold and another in the S-dual of the Klebanov-Strassler geometry \cite{ks}, also studied in \cite{kpv}.  Finally in the Appendices we describe the relation to the integral twisted homology of \cite{Andres}, which we extend to the non-$spin^c$ case, and we derive the quantization condition obeyed by integrals of RR field strengths on generalized cycles.

\section{D-brane networks and $\hat\partial$-homology}
\label{sec1}

In this section we will define a generalized boundary operator which acts on
submanifolds endowed with a field strength $\calf$.  This operator naturally
appears in the D-brane context and cannot be ignored in string theory if the
$H$ field is nontrivial.   While the ordinary boundary map $\partial$ takes a
$p$-submanifold $\Sigma$ to a $(p-1)$-submanifold $\partial\Sigma$, the image
of the generalized boundary operator $\hat{\partial}$ applied to a  $p$-submanifold $\Sigma$ may also contain a $(p-3)$-submanifold $\calc$.  We will identify $\calc$ with a Dirac monopole for $\calf$ on the worldvolume of a D$p$-brane wrapping $\Sigma$, or equivalently with the surface on which a D$(p-2)$-brane ends.  Therefore the  $\hat\partial$-homology classes will naturally be represented by sums of submanifolds of different degrees, all of which have the same parity.  We will refer to D-branes wrapping the elements of these complexes of submanifolds as D-brane networks.


\subsection{\Generalized\ chains}
\label{gchains}

A D-brane network is described by the embeddings of the constituent branes and their worldvolume gauge fields.  For a single D-brane, wrapping a submanifold $\Sigma$  with a field strength $\calf$, this data is summarized by a pair $(\Sigma,\calf)$, which is called a generalized submanifold  in the context of generalized complex geometry \cite{gualtieri}.  In this subsection we will slightly modify the original definition of  generalized submanifold and use it to introduce generalized chains, which will allow us to deal with D-brane networks.


Let $X$ denote the ten-dimensional spacetime.  We do not assume that its spatial slices are compact and we consider configurations in which both the NS $H$ field and RR field strengths may be nontrivial and properly quantized. If we use conventions in which $2\pi\sqrt{\alpha^\prime}=1$, the $H$ field satisfies the quantization condition
\bea\label{quantcond}
\int_{N_3} H\in \mathbb{Z}
\eea
for any compact three-cycle $N_3\subset X$.  The RR quantization condition is derived in Appendix B.

The worldvolume of a D$p$-brane extends along a $(p+1)$-submanifold $\Sigma\subset X$.  The worldvolume gauge theory of the D-brane contains a $U(1)$ gauge field, which may contain a codimension-three Dirac monopole with worldvolume $\calc$.  While we allow both $\Sigma$ and $\calc$ to be manifolds with boundary, the boundary of a magnetic monopole is always contained in the boundary of the host brane
\bea
\partial \calc\subset \partial\Sigma. \label{relcatena}
\eea
The condition (\ref{relcatena}) implies that $\calc$ is a $(p-2)$-cycle that represents a class $[\calc]$ in the relative homology group $\H_{p-2}(\Sigma,\partial\Sigma)$.  We will see momentarily that this class must exactly cancel the class Poincar\'e dual to the cohomology class of the pullback of the $H$-flux onto the D-brane worldvolume $\Sigma$,
\bea \label{calcH}
[\calc]+\textup{PD}_\Sigma([H|_\Sigma])=0.
\eea
We recall that, as $\Sigma$ may have a boundary, Poincar\'e-Lefschetz duality relates the relative homology of $\Sigma$ with respect to its boundary to the ordinary cohomology of~$\Sigma$.

We will denote the worldvolume field strength by $\calf_\calc$, so that the notation includes a choice of the magnetic source $\calc$. Now the position and gauge configuration of the D-brane is captured by the pair $(\Sigma,\calf_\calc)$, which we will still refer to as a \generalized\ submanifold. 
As is familiar in electrodynamics,  a magnetic monopole appears as a source term  in the Bianchi identities for the $U(1)$ field strength. In the present context, the gauge-invariant field strength $\calf_\calc$ can be locally split into the sum of a purely worldvolume $U(1)$ field strength  and the pullback of the $B$-field (locally characterized by $H=\d B$). Thus $\calf_\calc$ obeys the modified Bianchi identity 
\bea\label{modBI}
\d\calf_\calc=H|_\Sigma+\delta^{3}_\Sigma(\calc)\ .
\eea  

In general magnetic monopoles may consist of multiple components of various charges.  Thus we will generalize to this setting by allowing $\calc$ to be a codimension three chain, which is a weighted sum of submanifolds.  In (\ref{modBI}) the Dirac delta function  on a chain can be defined by asserting that it reduces to the usual Dirac delta function on a single submanifold and is linear.\footnote{More directly, if $\calc$ is $(p-3)$-chain inside $\Sigma$, $\delta^3_\Sigma(\calc)$ is defined by $\int_\Sigma \delta^3_\Sigma(\calc)\wedge\chi\equiv \int_\calc\chi$ for any $(p-3)$-form $\chi$ on $\Sigma$.}  The field strength $\calf_\calc$ is gauge-invariant and so the right hand side of (\ref{modBI}) must be exact.  This means that it represents the trivial cohomology class in $\H^3(\Sigma)$, and so $H|_\Sigma$ and $-\delta^{3}_\Sigma(\calc)$ must be cohomologous.  $\delta^3_\Sigma(\calc)$ is Poincar\'e dual to $\calc$, and so one recovers Eq.~(\ref{calcH}).  Condition (\ref{calcH}) constrains the possible configurations by corresponding tadpole conditions. Take for example $X=\mathbb{R}\times \calm$ and $\Sigma=\mathbb{R}\times\Gamma$,  where $\Gamma$ is a compact cycle in $\calm$. In this case, if $H|_\Gamma$ is exact, then there cannot be net magnetic monopole charge on $\Gamma$, even if  source terms in (\ref{modBI}) are not identically zero. When there are no source terms at all,  we will describe the D-brane configuration by the pair $(\Sigma,\calf)$.

We are now ready to define chains.  An even (odd) chain $(\mathfrak{S},\mathfrak{F})$ is a formal sum of \generalized\ submanifolds $(\Sigma^{(k)},\calf^{(k)}_{\calc^{(k)}})$ with integer coefficients in which the dimensions of all of the submanifolds are even (odd):
\bea
(\mathfrak{S},\mathfrak{F})=\sum_k n_k (\Sigma^{(k)},\calf^{(k)}_{\calc^{(k)}}) \quad ,\quad n_k\in\mathbb{Z}\ .
\eea


\subsection{\Generalized\ currents}


The generalized chains introduced in the previous section are naturally dual to polyforms on $X$ (formal sums of forms of different degrees) of definite parity. 
To this space of polyforms we can associate the  dual space of linear functionals defined on it, whose elements we call \generalized\ currents. In general, we may identify a generalized current $j$ on $X$ with a (distributional) polyform that we indicate with the same  symbol $j$ and is such that, for any (smooth compactly-supported) polyform $\phi$,
\bea
j(\phi)\equiv\int_X\langle \phi, j\rangle\label{cornonrit}
\eea
where $\langle \cdot, \cdot\rangle$ indicates the Mukai pairing\footnote{Given two polyforms $\phi_1$ and $\phi_2$ on a space $M$, the Mukai pairing is defined by the formula $\langle \phi_1, \phi_2\rangle=[\phi_1\wedge\sigma(\phi_2)]_{\text{top}}$, where $\sigma$ reverses the order of the indices of a form.} and only the top form is selected by the integration. In our context, we are interested in the particular class of generalized currents \cite{lucadef} associated to generalized submanifolds, and by linearity to generalized chains. 

Given a \generalized\ submanifold $(\Sigma,\calf_\calc)$, the associated \generalized\ current $j_{(\Sigma,\calf_\calc)}$ is defined by
\bea
j_{(\Sigma,\calf_\calc)}(\phi)= \int_X\langle \phi, j_{(\Sigma,\calf_\calc)}\rangle \equiv 
\int_{\Sigma}\phi|_\Sigma\wedge e^{\calf_\calc}. \label{PD}
\eea
Note that these functionals are well defined even if $\calf_\calc$ is singular at $\calc$. Indeed, consider a tubular neighborhood of $\calc$ which is a $D^3$-bundle over $\calc$.  This bundle can be foliated in $S^2$ bundles over $\calc$. For any sufficiently small two-sphere $S^2$ linking $\calc$, $\int_{S^2}\calf_\calc$ is finite and (almost) constant  by Gauss' Law. Therefore the monopole source does not make  the above integral over the full $D^3$-bundle, and thus over the full $\Sigma$, diverge.


The definition (\ref{PD}) of the current associated to a generalized submanifold is suggested by the D-brane effective action. Since in general this action includes gravitational corrections, one would naturally be led to a slightly different  definition 
\bea\label{corrcorr}
j^{\text{grav}}_{(\Sigma,\calf_\calc)}=j_{(\Sigma,\calf_\calc)}\wedge \sqrt{\hat{A}(TX)}\ ,
\eea
where $\hat{A}(T_X)$ is the A-roof genus of the background tangent bundle.  
The leading terms in the A-roof genus are $\hat{A}=1+p_1/24+\dots$, where $p_1$ is a 4-form called the first Pontrjagin class of the tangent bundle and $\dots$ are higher degree forms.  The fact that the degree 0 term is equal to one and the degree 2 term vanishes means that the gravitational corrections only appear at degree 4, like the corrections from gauge instantons in the nonabelian case.  As in that case, these corrections will not change the classification of charges, but merely shift the charges which are already there.  In general these lead to interesting physical effects, like half-integer shifts in the quantization condition and gravitational corrections to the calibration formulas.  In the present paper we will ignore these corrections, that may be nevertheless reintroduced by using the corrected generalized currents $(\ref{corrcorr})$ instead of the uncorrected ones.


\subsection{The generalized boundary operator} \label{ritorta}

What have we gained by reexpressing our \generalized\ submanifolds in terms of \generalized\ currents?  Our goal is to construct a homology theory that classifies equivalence classes of continuously connected \generalized\ chains that can be wrapped by D-brane networks.  To do this we must choose a generalized boundary operator.  A generalized boundary operator that acts on \generalized\ submanifolds with sources has not yet appeared in the physics literature.  

On the other hand, there is a twisted exterior derivative
\bea
\d_H\equiv \d+H\wedge
\eea
that acts on polyforms.  RR field strengths are $\d_H$-closed and can be locally written as $F=\d_H C$, where we assemble the RR gauge fields and the  field strengths into the polyforms $C$ and $F$ respectively, in such a way that an infinitesimal RR gauge transformation is given by $\delta C=\d_H\lambda$, where $\lambda$ is any polyform (of degree opposite to $C$).  As  RR fluxes are sourced by D-branes, we are led naturally to the proposal that the generalized boundary operator whose homology classifies D-branes is just the dual of the twisted exterior derivative.  In subsection \ref{omrit} we will see that this proposal implies that consistent D-branes always wrap generalized chains which are closed with respect to this generalized boundary operator and are thus generalized cycles.  In sec. \ref{genhom} we will show that this guess successfully describes physical processes such as branes dissolving in branes and MMS instantons \cite{mms}, as the original and final configurations always differ by a generalized boundary.  

We will now use the definition (\ref{PD}) to calculate the action of the generalized boundary operator on an arbitrary \generalized\ submanifold $(\Sigma,\calf_\calc)$ as the operator dual to the action of the $\d_H$-differential on the  \generalized\ current $j_{(\Sigma,\calf_\calc)}$. 
Using Stokes' theorem and the modified Bianchi identity (\ref{modBI}), for any polyform $\phi$ on $X$ one finds
\bea \label{diffrit}
\int_{X} \langle \phi, \d_H j_{(\Sigma,\calf_\calc)}\rangle&\equiv&\int_{ X} \langle \d_H\phi, j_{(\Sigma,\calf_\calc)}\rangle=\int_{\Sigma} \d_H\phi|_{\Sigma}\wedge e^{\calf_\calc}\cr &=&\int_{\partial\Sigma}\phi|_{\partial\Sigma}\wedge e^{\calf_\calc|_{\partial\Sigma}}-\int_{\calc}\phi|_{\calc}\wedge e^{\calf_\calc|_{\calc}}\cr
& =&\int_{ X} \langle \phi, j_{(\partial\Sigma,\calf_\calc|_{\partial\Sigma})}- j_{(\calc,\calf_\calc|_\calc)}\rangle \ .
\eea
The polyform $\phi$ is arbitrary and so one can read the action of the twisted differential on $j_{(\Sigma,\calf_\calc)}$ off of Eq.~(\ref{diffrit})
\bea\label{diffj}
\d_Hj_{(\Sigma,\calf_\calc)}=j_{(\partial\Sigma,\calf_\calc|_{\partial\Sigma})}- j_{(\calc,\calf_\calc|_\calc)}\ .
\eea
Notice that, while $\d\calf_\calc$ is singular at $\calc$, $\calf_\calc|_\calc$
is nevertheless well defined. Indeed, $\delta^{3}_\Sigma(\calc)$ has all legs (co)normal to $\calc$ and so is killed by the restriction to $\calc$, therefore (\ref{modBI}) reduces to
\bea
\d \calf_\calc|_\calc=H|_\calc
\eea
which means that the field strength $\calf|_\calc$ on $\calc$ has no magnetic sources. 

If $\calc$ has a nontrivial boundary $\partial\calc\subset \partial\Sigma$, then the definition $H=\d B$ combined with the definition of $\calc$ as the magnetic monopole worldvolume implies\footnote{The relation (\ref{restrBI}) is valid even in the degenerate case in which not only $\partial\calc$ but  also $\calc$ itself is contained in $\partial\Sigma$, as can be seen by  `regularizing' it by slightly deforming $\calc$ in such a way that $\calc-\partial\calc\subset \Sigma-\partial\Sigma$.}
\bea\label{restrBI}
\d\calf_{\calc}|_{\partial\Sigma}=H|_{\partial\Sigma}-\delta^3_{\partial\Sigma}(\partial\calc)\ ,
\eea
where we have used the relation
\bea
\delta^3_{\Sigma}(\calc)|_{\partial\Sigma}=-\delta^3_{\partial\Sigma}(\partial\calc)\ .
\eea
The relative minus sign is due to the fact that if we use the natural orientations of $\partial\Sigma$ and $\calc$ induced by their inclusions into $\Sigma$, the orientations on $\partial\calc$ induced by its inclusions in $\partial\Sigma$ and in $\calc$ disagree.

Finally we are ready to define the generalized boundary operator $\hat\partial$ by imposing
\bea
\d_Hj_{(\Sigma,\calf_\calc)}=j_{\hat\partial(\Sigma,\calf_\calc)}\ .
\eea
Eq.~(\ref{diffj}) now yields the generalized boundary operator $\hat\partial$ on \generalized\ chains of definite dimension on the constituent generalized submanifolds 
\bea\label{gboundary}
\hat\partial(\Sigma,\calf_\calc)=(\partial\Sigma,\calf_\calc|_{\partial\Sigma})- (\calc,\calf_\calc|_\calc)\ ,
\eea 
which extends to arbitrary chains by linearity.
We will refer to \generalized\ chains in the kernel of $\hat\partial$ as generalized cycles and \generalized\ chains in the image of $\hat\partial$ as generalized boundaries. Using $\d \calf_\calc|_\calc=H|_\calc$ and the relation (\ref{restrBI}), one finds that that $\hat\partial^2=0$ and thus $\hat\partial$ may be a suitable boundary operator for a homology theory.  Of course this works because $H\wedge H=0$, which is true at the level of differential forms but may fail in the full integral homology due to possible torsional effects, possibly leading to interesting physical effects.   We will comment more on this in Appendix~\ref{torsion}.


\subsection{Generalized cycles} \label{omrit}

The nilpotency of $\hat\partial$ implies that its image is a subgroup of its kernel and we define the even and odd $\hat\partial$-homology groups to be the quotient of the kernel of $\hat\partial$ by its image
\bea \label{omritdef}
\hat\H_{\textup{\scriptsize{even}}}=\frac{\Ker(\hat\partial:\Upsilon_{\textup{\scriptsize{even}}}\longrightarrow \Upsilon_{\textup{\scriptsize{odd}}})}{\Im(\hat\partial:\Upsilon_{\textup{\scriptsize{odd}}}\longrightarrow \Upsilon_{\textup{\scriptsize{even}}})}\hsp
\hat\H_{\textup{\scriptsize{odd}}}=\frac{\Ker(\hat\partial:\Upsilon_{\textup{\scriptsize{odd}}}\longrightarrow \Upsilon_{\textup{\scriptsize{even}}})}{\Im(\hat\partial:\Upsilon_{\textup{\scriptsize{even}}}\longrightarrow \Upsilon_{\textup{\scriptsize{odd}}})}
\eea
where $\Upsilon_{\textup{\scriptsize{odd}}}$ and $\Upsilon_{\textup{\scriptsize{even}}}$ are the groups of odd and even generalized chains respectively.


From (\ref{gboundary}) we see that a  \generalized\ chain consisting of a single generalized submanifold $(\Sigma,\calf_\calc)$  is a generalized cycle only if $\Sigma$ is a cycle with respect to the ordinary boundary operator $\partial$ and $\calc$ is empty. However, if we allow chains in which the underlying submanifolds $\Sigma^{(k)}$ have different dimensions, we find more interesting generalized cycles.  For example if $\Sigma$ and $\Sigma\p$ are of dimension $p$ and $p-2$ respectively and if $\Sigma$ supports a nontrivial $H$ flux whose cohomology class  is dual to homology class defined by $\calc$ then
\bea\label{simplegc}
(\Sigma,\calf_\calc)+(\Sigma^\prime,\calf^\prime)
\eea
is a generalized cycle if $\Sigma$ is a $p$-cycle and if furthermore $\partial\Sigma^\prime=\calc$ and $\calf_\calc|_{\calc}=\calf^\prime|_{\partial\Sigma^\prime}$. 
More generally we may consider generalized cycles consisting of formal sums of several \generalized\ chains of different definite dimensions\footnote{We use  the symbol $(\mathfrak{S},\mathfrak{F})$ to denote a \generalized\ chain, while  $(\mathfrak{C},\mathfrak{F})$ will denote a generalized cycle.}
\bea\label{gc2}
(\mathfrak{C},\mathfrak{F})=(\Sigma^{(1)},\calf^{(1)}_{\calc^{(1)}})+(\Sigma^{(2)},\calf^{(2)}_{\calc^{(2)}})+(\Sigma^{(3)},\calf^{(3)}_{\calc^{(3)}})+\ldots\ ,
\eea 
where $\Sigma^{(1)}$ is a cycle and, for $k\geq 2$,  $\Sigma^{(k)}$ are chains such that $\partial\Sigma^{(k)}=\calc^{(k-1)}$ and $\calf^{(k-1)}_{\calc^{(k-1)}}|_{\calc^{(k-1)}}=\calf^{(k)}_{\calc^{(k)}}|_{\partial\Sigma^{(k)}}$. By construction if $(\mathfrak{C},\mathfrak{F})$ is a generalized cycle,  then the corresponding dual \generalized\ current
\bea
j_{(\mathfrak{C},\mathfrak{F})}=\sum_k j_{(\Sigma^{(k)},\calf^{(k)}_{\calc^{(k)}})}
\eea
is $\d_H$-closed.

The formal sum of different \generalized\ submanifolds in a \generalized\ chain has a clear physical interpretation in terms of networks that consist of several D-branes of different dimensions, where some D$p$-branes may end on D$(p+2)$-branes. D$p$-branes ending on yet higher dimensional branes, as are present for example in the SU(3) WZW model studied in \cite{mms}, exist because of the quantum Freed-Witten anomaly.  In the generalized lens space compactifications of \cite{hisham} such configurations are the result of an anomaly on branes wrapping certain cycles that cannot be represented by submanifolds. Both of these effects are missed in our present formalism, which is insensitive to quantum corrections.  

The gauge invariance of the Chern-Simons terms in the D-brane actions requires that all networks consist of D-branes wrapping  a total generalized cycle.  Consider a D-brane network wrapping a generalized chain $(\mathfrak{S},\mathfrak{F})$.  The total Chern-Simons term is\footnote{To simplify notation, here and in the following expressions involving D-brane actions and energies we omit an overall factor $2\pi$ given by the D-brane's tension in units $2\pi\sqrt{\alpha^\prime}=1$.}
\bea
S_{\text{CS}}=
\int_{ X}\langle C,j_{(\mathfrak{S},\mathfrak{F})}\rangle\ ,
\eea    
where we have used the total RR gauge connection $C=\sum_{\text{even/odd}} C_{(k)}$. Under an infinitesimal RR gauge transformation $\delta_\lambda C=\d_H\lambda$ and thus
\bea
\delta_\lambda S_{\text{CS}}=\int_{X}\langle \lambda,\d_Hj_{(\mathfrak{S},\mathfrak{F})}\rangle=\int_{X}\langle \lambda,j_{\hat\partial(\mathfrak{S},\mathfrak{F})}\rangle ,
\eea
implying that the Chern-Simons term is invariant only if  
$(\mathfrak{S},\mathfrak{F})$ is a generalized cycle. 

When this condition is satisfied one finds the expected possible consistent configurations. For simplicity consider a generalized chain which is the sum of two generalized submanifolds $(\Sigma,\calf_\calc)$ and $(\Sigma^\prime,\calf^\prime_{\calc^\prime})$.  If $\Sigma$ has no boundary then there are two possibilities.  First $\Sigma^\prime$ may also be a cycle. In this case, generically, neither brane carries net monopole charge $\calc=\calc^\prime=0$ and the two generalized cycles of definite dimension are separately allowed D-brane configurations. However, if the two cycles have the same dimension, we can also consider the case when $\Sigma$ and $\Sigma^\prime$ intersect at $\calc=-\calc^\prime$ and $\calf_\calc|_\calc=-\calf^\prime_{\calc^\prime}|_{\calc^\prime}$, so that they have opposite magnetic sources that can annihilate each other.  If on the other hand $\Sigma$ and $-\Sigma^\prime$ are homologous then $\calc$ and $-\calc^\prime$ are homologous and one finds a meson-like bound state in which the $p$-branes wrapping $\Sigma$ and $-\Sigma^\prime$ are attached by a $(p-2)$-brane extending from $\calc$ to $\calc^\prime$.  Another possibility is that the D-branes wrap a generalized cycle of the kind described in (\ref{simplegc}), where
the D-branes wrapping $(\Sigma^\prime,\calf^\prime)$  end on the D-branes wrapping $(\Sigma,\calf_\calc)$, playing the role of worldvolume magnetic monopoles wrapped on the submanifold $\calc$.  

We thus see that the  \generalized\ chains introduced above and  the generalized boundary operator $\hat\partial$ acting on them  as defined in (\ref{gboundary}) allow one to characterize in a concise language the general consistent  D-brane networks simply in terms of generalized cycles. The natural next step is  to provide a physical interpretation for the homological equivalence  defined by $\hat\partial$.  This will be the subject of the next section.

\section{Generalized homologous configurations,  MMS instantons and dissolving branes} 
\label{genhom}

We have seen that generalized cycles are precisely those chains that can be wrapped by consistent D-brane networks.  In this section we will interpret generalized boundaries, which are all identified with the zero element in $\hat\partial$-homology.  We will argue that a D-brane network can decay if and only if the generalized cycle that it wraps is a generalized boundary.  

There are several distinct decay processes encoded in the $\hat\partial$-homology formalism.  A D-brane may decay if it wraps a contractible cycle and carries no lower-dimensional D-brane charges by simply collapsing.  It may collapse via a series of worldvolume topology-changing transitions even if it wraps a noncontractible $p$-cycle which is the boundary of a $(p+1)$-cycle by sweeping out the $(p+1)$-cycle.  Also it may annihilate an anti-brane which has all opposite charges.  It may dissolve into the gauge field of a higher-dimensional D-brane.  Finally a D$p$-brane wrapping a nontrivial cycle which is orthogonal to a nontrivial $H$ flux on a compact 3-cycle may grow into a D$(p+2)$-brane, in a Myers dielectric configuration \cite{Myers}, sweep out the cycle and then shrink into oblivion.  These last decay processes, known as MMS instantons, were used in \cite{mms} to argue that D-branes are classified by twisted K-theory.\footnote{Here we neglect  a possible quantum correction, the third Stiefel-Whitney class, on the worldvolume of the D$(p+2)$-branes and so arrive instead at $\hat\partial$-homology. See Appendix~\ref{torsion} for a discussion of this point.
}

\begin{figure}[ht]
\begin{center}
\leavevmode
\epsfxsize 12   cm
\epsffile{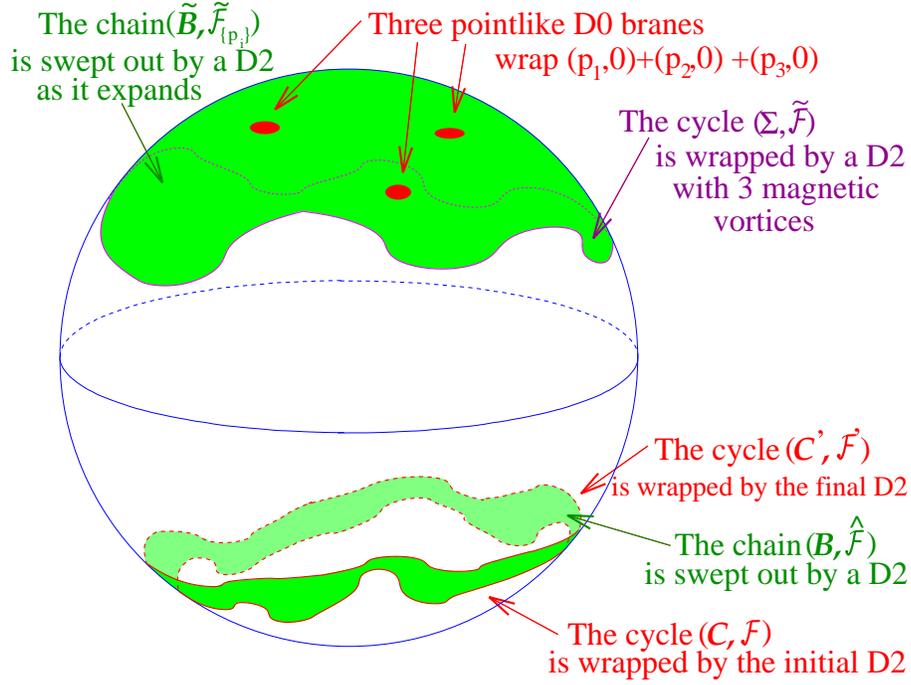}    
\end{center} 
\caption{\small{Here we see two dynamical processes in which a network transforms into a different network in the same $\hat\partial$-homology class.  In the southern hemisphere a brane wrapping the cycle $C$ with gauge bundle $\calf$ moves, sweeping out the chain $B$.  As it moves the gauge bundle $\hat{\calf}$ also changes, although it always satisfies $\d\hat{\calf}=H|_B$.  Finally it arrives at the cycle $C\p$ where the gauge bundle is $\hat{\calf}|_{C\p}=\calf\p$.  In the northern hemisphere 3 D0-branes swell to form a D2-brane wrapping a trivial cycle but with 3 worldvolume magnetic vortices.  They sweep out the chain $(\tilde{B},\tilde{\calf}_{\{ p_i\}})$, whose worldvolume flux $\tilde{\calf}_{\{ p_i\}}$ is sourced at the positions $p_i$ of the original D0-branes.}}
\label{myersfig} 
\end{figure}

More explicitly, consider two generalized cycles, $(\mathfrak{C},\mathfrak{F})$ and $(\frakc^\prime,\frakf^\prime)$ of the same parity.  These two cycles are said to be homologous if there exists a chain $(\fraks,\hat\frakf)$ of the opposite parity such that
\bea
\hat\partial(\fraks,\hat\frakf)=(\frakc^\prime,\frakf^\prime)-(\mathfrak{C},\mathfrak{F}).
\eea
In this case the difference $(\frakc^\prime,\frakf^\prime)-(\mathfrak{C},\mathfrak{F})$\ is in the image of $\hat\partial$ and so is a generalized boundary.  If the difference between two D-brane configurations is a boundary then they represent the same $\hat\partial$-homology class.  We claim that in this case a physical process may occur in which the network $(\mathfrak{C},\mathfrak{F})$ becomes the network $(\frakc^\prime,\frakf^\prime)$.  This physical process is just the nucleation of the unstable network that sweeps out $(\fraks,\hat\frakf)$.  At this point we have not considered the dynamics of the system, and so we do not know whether such a transition will happen or if it is even energetically favorable.  The difference in energy between the initial and final state of such a process will be considered in Sec.~\ref{calisec} where we endow the backgrounds where the networks live with calibrations.  In this section we will instead show that in examples these processes are in line with physical expectations.

For concreteness let us isolate the time direction $\mathbb{R}$ in the ten-dimensional spacetime $X=\mathbb{R}\times \calm$ and restrict our attention to  two generalized cycles of definite dimension (and thus with no magnetic sources) $(\Sigma,\calf)$ and  $(\Sigma^\prime,\calf^\prime)$ such that $\Sigma=\mathbb{R}\times \Gamma$, $\Sigma^\prime=\mathbb{R}\times \Gamma^\prime$ and the worldvolume field strengths live only on the internal cycles $\Gamma,\Gamma^\prime\subset \calm$.\footnote{We would like to stress that our definition of $\hat\partial$ given in (\ref{gboundary}) was derived for D-branes in a ten-dimensional background $X$. If we split $X$ and the generalized chains into direct products of lower dimensional objects (like in the present examples), the restriction of the definition of $\hat\partial$ to the different subsectors  could require some sign changes to take into account the orientation of the total generalized chains.}
First we consider the case in which the two cycles $\Gamma$ and $\Gamma^\prime$ are of the same dimension and are homologous with respect to the ordinary boundary operator $\partial$, as depicted in the southern hemisphere of Fig.~\ref{myersfig}. In this case our generalized cycles $(\Sigma,\calf)$  and $(\Sigma^\prime,\calf^\prime)$ are $\hat\partial$-homologous if there exists a \generalized\ chain $(\calb,\hat\calf)$ on $\calm$ such that
\bea
\partial\calb=\Gamma^\prime-\Gamma\hsp\hat\calf|_\Gamma=\calf\hsp\hat\calf|_{\Gamma^\prime}=\calf^\prime\ .
\eea
This means that each of the generalized cycles can be continuously deformed to the other and so they are equivalent in both ordinary and in $\hat\partial$-homology.  This case agrees with the homology relation used in \cite{luca1,luca2,lucadef,jarahrev}, where possible source terms in (\ref{modBI}) have not been considered and thus each brane can wrap only proper cycles.

\begin{figure}[ht]
\begin{center}
\leavevmode
\epsfxsize 12   cm
\epsffile{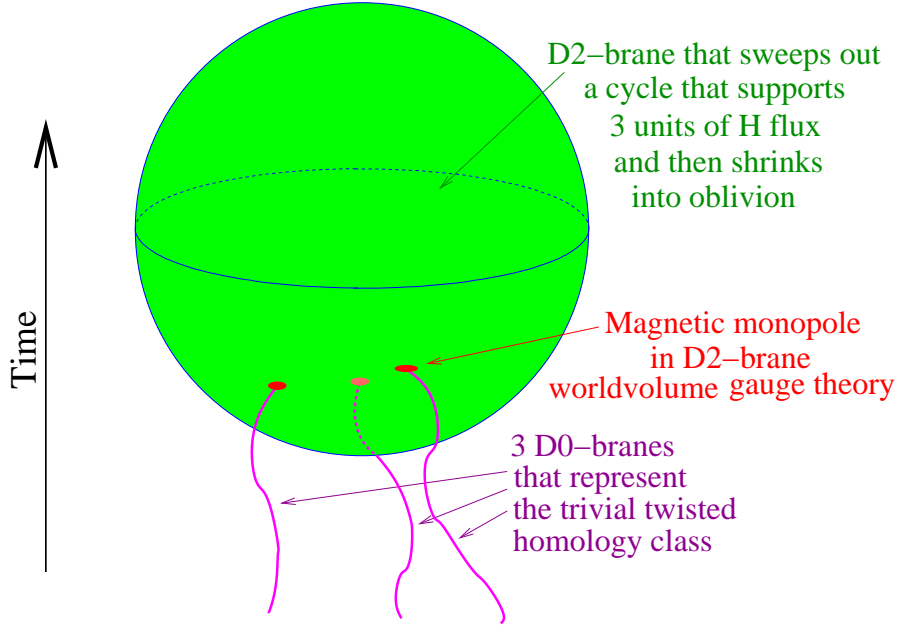}    
\end{center} 
\caption{\small{3 D0-branes represent a trivial homology class.  They therefore can decay via an MMS instanton, which is a D2-brane that sweeps out a 3-cycle $\Sigma$ such that the endpoints of the D0-branes are Poincar\'e dual in $\Sigma$ to the pullback of the $H$ flux.  The D0's inflate into a spherical D2-brane via the Myers effect, sweep out the 3-cycle and then disappear as the sphere shrinks to nothing at the north pole.  This process creates a residual RR 6-form field strength.}}
\label{MMSfig} 
\end{figure}

Now we will consider a more interesting example in which the two cycles are chosen to have dimensions
\bea
\dim\Sigma^\prime=\dim\Sigma +2
\eea
as is depicted in Fig.~\ref{MMSfig}.  In this case $(\Sigma,\calf_\calc)$ and  $(\Sigma^\prime,\calf^\prime_{\calc^\prime})$ are $\hat\partial$-homologous if some interpolating chain $(\calb,\hat\calf_{\hat\calc})$ in $\calm$ (i.e. a generalized chain $(\mathbb{R}\times \calb, \hat\calf_{\mathbb{R}\times\hat\calc})$ in $X$) satisfies
\bea
\partial\calb=-\Gamma^\prime\quad\quad\text{and}\quad\quad\hat\calc=\Gamma\ ,
\eea
so that $\hat\calf_{\hat\calc}$ is sourced by a delta function with support on $\Gamma$. As usual, the  consistency condition (\ref{calcH}) must be satisfied. In particular $(\Sigma,\calf_\calc)$ can be  trivial in $\hat\partial$-homology even if $\Gamma$ is a nontrivial cycle in ordinary homology but there exists a \generalized\ chain $(\calb,\hat\calf_{\hat\calc})$ in $\calm$ such that $\calb$ is a cycle and $\hat\calc=\Gamma$. The Euclidean D-brane instanton wrapping $(t_0\times\calb,\hat\calf_{t_0\times\hat\calc})$, at some time $t_0\in\mathbb{R}$, is an extension of the MMS D-brane instantons  \cite{mms}, and the total consistent D-brane configuration undergoing the instanton transition is given by the generalized cycle
\bea
(\frakc,\frakf)=((-\infty,t_0]\times\Gamma^\prime,\calf^\prime)+([t_0,+\infty)\times \Gamma,\calf)+(t_0\times\calb,\hat\calf_{t_0\times\hat\calc})\ .
\eea  

We have explained that trivial $\hat\partial$-homology classes correspond to potentially unstable D-brane networks. Adding a constant charge to the initial and final state thus implies that generalized cycles which represent the same $\hat\partial$-homology class are related by dynamical processes.  One dynamical process which can be described in this way is the dissolving of a stack of $n$ D$p$-branes in a D$(p+2)$-brane.  In the worldvolume theory of the D$(p+2)$-brane, once the D$p$-branes are dissolved their RR charges are carried by a charge $n$ magnetic vortex with respect to the D$(p+2)$-brane's worldvolume $U(1)$ gauge field.  We will now show that the generalized cycle which describes the D$p$-branes dissolved in the D$(p+2)$ is homologous to that in which they are not dissolved and the D$(p+2)$-brane's worldvolume gauge field is trivial.  

We suppose for simplicity that there is no $H$-flux
and begin with a D$(p+2)$-brane wrapping the $(p+3)$-cycle $\Sigma^{(1)}$ and $n$ D$p$-branes wrapping the $(p+1)$-cycle  $\Sigma^{(2)}\subset\Sigma^{(1)}$. Our starting D-brane network is characterized by the generalized cycle
\bea
(\Sigma^{(1)},0)+n(\Sigma^{(2)},0)\ ,
\eea 
where no branes are dissolved and all worldvolume fluxes are trivial.

Now consider a $(p+3)$-cycle $\Sigma^\prime$ which is homotopic to $\Sigma^{(1)}$, so that  a $(p+4)$-chain $\calb$ exists satisfying $\partial\calb=\Sigma^\prime-\Sigma^{(1)}$ \footnote{We may even  consider $\Sigma^\prime=\Sigma^{(1)}$, but it is useful to regularize this kind of configurations by choosing a  $\Sigma^\prime$ which is a slight deformation of $\Sigma^{(1)}$.}. We can construct a \generalized\ chain $(\calb,\hat\calf_{\hat\calc})$  describing the deformation of $(\Sigma^{(1)},0)+n(\Sigma^{(2)},0)$ to a generalized cycle $(\Sigma^\prime,\calf)$ if
\bea \label{dhfunziona}
\hat\partial(\calb,\hat\calf_{\hat\calc})=(\Sigma^\prime,\calf)-(\Sigma^{(1)},0)-n(\Sigma^{(2)},0)\ ,
\eea
which implies
\bea
\hat\calc=n\Sigma^{(2)}\quad,\quad \hat\calf_{\hat\calc}|_{\Sigma^\prime}=\calf\quad,\quad\hat\calf_{\hat\calc}|_{\Sigma^{(1)}}=\hat\calf_{\hat\calc}|_{\Sigma^{(2)}}=0\ .
\eea
The fact that the D$p$-branes are magnetic monopoles in the worldvolume of the D$(p+2)$-brane sweeping out $\calb$ implies that
\bea
\d\hat\calf_{\hat\calc}=n\delta^3_\calb(\Sigma^{(2)})
\eea
which integrated over each cross section yields 
\bea
\int_{N^\prime_2}\calf=\int_{B_3}\d\hat{\calf}_{\hat\calc}=n\hsp \partial B_3=N_2\p-N_2
\eea  
where $B_3$ is a 3-chain which relates two homotopic cycles $N_2\subset\Sigma^{(1)}$
and $N_2^\prime\subset\Sigma^\prime$, where $N_2$  has intersection number one with $\Sigma^{(2)}\subset \Sigma^{(1)}$. We have thus arrived at the usual condition that we must have $n$ quanta of worldvolume flux on the final D$(p+2)$-brane, which wraps $\Sigma^\prime$. These units of flux couple to the RR $(p+1)$-form gauge connection just like the dissolved $n$ D$p$-branes, and so D$p$-brane charge is conserved by this process.  The case in which 3 D0-branes dissolve in a contractible D2 is depicted in the northern hemisphere of Fig.~\ref{myersfig}.

More generally we may start with a configuration in which the D$p$-branes wrap a $(p+1)$-cycle $\Sigma^{\prime(2)}$ which is not included in  $\Sigma^{(1)}$ but can be continuously connected to $\Sigma^{(2)}\subset \Sigma^{(1)}$ by a $(p+2)$-chain $\cala$ such that $\partial\cala=\Sigma^{\prime(2)}-\Sigma^{(2)}$.  In this case the total \generalized\ chain describing the deformation to $(\Sigma^\prime,\calf)$ is
\bea
n(\cala,0)+(\calb,\hat\calf_{\hat\calc})\ 
\eea
and so it would be possible to move the D$p$-branes to the D$(p+2)$-brane and then dissolve them inside. 



From this discussion, it seems that the homology defined  by $\hat\partial$ gives a natural and powerful topological characterization of D-brane charges.  We will see that it is not necessarily true that a contractible generalized cycle has trivial charge, and thus one cannot naively identify the $\hat\partial$-homology of a certain D-brane configuration with its RR charge.  

\section{Supersymmetry and \generalized\ calibrations} \label{calisec}


Until now we have focused on the topological properties which characterize consistent D-brane configurations and their equivalence relations, without worrying about whether these configurations are in fact solutions of the equations of motion and thus contribute in a significant way to the path integral. In order to address this problem, we have to be a bit more specific about the supergravity background. We are interested in static type II configurations in a ten-dimensional spacetime of the form $\mathbb{R}\times \calm$ where $\mathbb{R}$ is the time direction. 
The most natural question is whether there exist static D-brane networks and, if so, whether they are stable. In this case,  the relevant  information can be obtained from the  energy of the configuration under consideration. Consider a consistent D-brane network wrapping the time direction and an internal generalized cycle $(\frakc,\frakf)=\sum_k (\Gamma^{(k)},\calf^{(k)}_{\calc^{(k)}})$ in $\calm$.  Its energy per unit volume  is given by\footnote{In this paper we do not consider cases in which the D-brane configurations admit only a pure non-commutative description.}
 \bea\label{energy}
E(\frakc,\frakf)\equiv \sum_k \int_{\Gamma^{(k)}}\cale(\Gamma^{(k)},\calf^{(k)}_{\calc^{(k)}})
\eea
where $\cale(\Gamma,\calf_\calc)$ denotes the energy density of the generalized submanifold $(\Gamma,\calf_\calc)$. Note that we expect the magnetic source at $\calc$ to give an infinite contribution to the energy. We will come back to this point soon.

Now it is clear that a consistent D-brane network wrapping $(\frakc,\frakf)$ is stable (or at least degenerate) only if every other homologous generalized cycle  $(\frakc^\prime,\frakf^\prime)$ satisfies
\bea
E(\frakc,\frakf)\leq E(\frakc^\prime,\frakf^\prime)\ .
\eea
Notice that, using this definition of stability, we are considering at the same time both the classical stability under continuous deformation and the stability against decays driven by the nucleation or annihilation of D-branes, or by the dissolution of D$p$-branes in D$(p+2)$-branes. However, we are completely neglecting other possible sources of instability (or metastability), caused for example by the nucleation and annihilation of NS5-branes, as was studied in \cite{kpv}, although that process is S-dual to an MMS process that is covered by our formalism. These and other possible phenomena should be automatically included in a description which is covariant under the non-perturbative dualities of type II theories, but even in this case it would face the issue of the applicability of the tree-level low energy description which is the starting point of our discussion.   In addition one would need to include the energy contribution of the bulk fields, which changes during an MMS process.  We are also ignoring processes in which D$p$-branes dissolve in stacks of D$(p+4)$-branes where they become instantons in their worldvolume nonabelian gauge theories, to cover these cases one would need to use the nonabelian Born-Infeld energy in (\ref{energy}).

Experience teaches us that the fact that supersymmetric D-branes are stable can be encoded in the presence of background calibrating forms that `calibrate' the supersymmetric branes. In order to take into account a possibly nontrivial background with worldvolume fluxes on the branes, we use the definition of generalized calibration used in \cite{luca1} (see also \cite{paulk}). In a static background $\mathbb{R}\times \calm$, where $\mathbb{R}$ is a time-like Killing direction parametrized by $t$, a \generalized\ calibration is a $\d_H$-closed polyform $\omega$  of definite parity (even in IIA and odd in IIB) on $X$ such that, for any D$p$-brane wrapping any static \generalized\ submanifold $(\mathbb{R}\times\Gamma,\calf)$ (with $\calf$ completely on $\Gamma$)
\bea\label{calineq}
[\omega|_\Gamma\wedge e^\calf]_{\text{top}}\leq \cale(\Gamma,\calf)\ ,
\eea  
where $[\ ]_{\text{top}}$ means that one takes the coefficient of the top form component.  A D-brane is said to be calibrated if the inequality (\ref{calineq}) is saturated at each point on its worldvolume.  Notice that this definition is completely local, which is why we have omitted possible source terms for $\calf$.  Let us also recall that we can single out the RR contribution to the generalized calibration in such a way that we can use a different kind of generalized calibration $\hat\omega=\omega+\iota_{\partial_t}C$ that  is gauge invariant under RR gauge transformations. $\hat\omega$ minimizes only the DBI energy density  i.e. \bea\label{calineq2}
[\hat\omega|_\Gamma\wedge e^\calf]_{\text{top}}\leq \cale_{\text{DBI}}(\Gamma,\calf)\quad,\quad \quad \cale_{\text{DBI}}(\Gamma,\calf)=e^{-\Phi}\sqrt{-\det(g|_{\mathbb{R}\times \Gamma}+\calf)}\,\d^{\rm top}\sigma\ ,
\eea
and satisfies the differential condition\footnote{Eq.~(\ref{diffcond2}) implies that the electric component $\iota_{\partial_t}F$ of the RR field-strength is $\d_H$-exact. We expect this to be valid in general, provided that there are no subtleties associated to special topological properties of space-time and we can safely identify the energy-density of the system with the standard DBI+CS energy density.}
\bea\label{diffcond2}
\d_H\hat\omega=\iota_{\partial_t}F.
\eea

We are thus led to restrict our attention to D-calibrated backgrounds. The explicit example  provided by the $\caln$=1 vacua considered in \cite{luca1} will be discussed in section~\ref{fluxsec}.  Now we will demonstrate a general feature of \generalized\ calibrations. Historically they were introduced to characterize  D-branes wrapping generalized cycles of definite dimension which are energy minimizing under continuous deformations.  However the above definition of \generalized\ calibrations characterizes D-brane networks which are energy minimizing not only under continuous deformations, but also under processes involving  nucleation, creation  or annihilation of D-branes.  Consider for example a D-brane network wrapping a static calibrated generalized cycle $(\frakc,\frakf)=\sum_k (\mathbb{R}\times\Gamma^{(k)},\calf^{(k)}_{\calc^{(k)}})$. Consider also a homologous static generalized cycle $(\frakc^\prime,\frakf^\prime)=\sum_r (\mathbb{R}\times\Gamma^{\prime(r)},\calf^{\prime(r)}_{\calc^{\prime(r)}})$.  Then there exists a \generalized\ chain $(\fraks,\hat\frakf)$ such that
\bea
\hat\partial (\fraks,\hat\frakf)=(\frakc^\prime,\frakf^\prime)- (\frakc,\frakf).
\eea
Now we can show that calibrated networks really do minimize energy
 \bea\label{defnetw}
 E(\frakc,\frakf)&=& \sum_k \int_{\Gamma^{(k)}}\omega|_{\Gamma^{(k)}}\wedge e^{\calf^{(k)}_{\calc^{(k)}}}=\int_\calm\langle \omega,j_{(\frakc,\frakf)}\rangle\cr
  &=&\int_\calm\langle \omega,j_{(\frakc^\prime,\frakf^\prime)}\rangle -\int_\calm\langle \omega,j_{\hat\partial (\fraks,\hat\frakf)}\rangle\cr
  &=& \sum_r \int_{\Gamma^{\prime(r)}}\omega|_{\Sigma^{\prime(r)}}\wedge e^{\calf^{\prime(r)}_{\calc^{\prime(r)}}}\leq E(\frakc^\prime,\frakf^\prime)\ ,
 \eea
where in going from the second to the third line we have used the fact that
\bea \label{zero}
\int_\calm\langle \omega,j_{\hat\partial (\fraks,\hat\frakf)}\rangle=\int_\calm\langle \omega,\d_H j_{(\fraks,\hat\frakf)}\rangle=-\int_\calm\langle \d_H\omega,j_{(\fraks,\hat\frakf)}\rangle=0\ .
\eea
The final equality in (\ref{zero}) is a consequence of the fact that $\omega$ is $\d_H$-closed.
\begin{figure}[ht]
\begin{center}
\leavevmode
\epsfxsize 12   cm
\epsffile{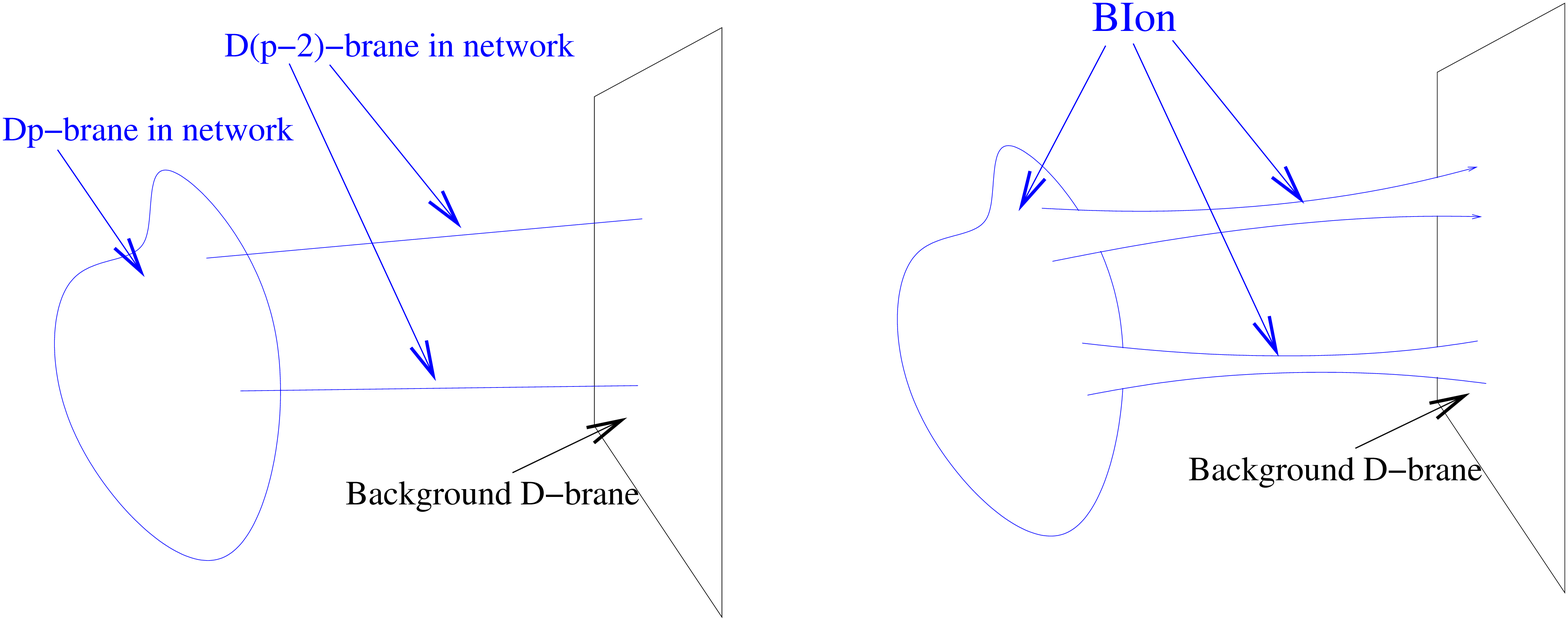}    
\end{center} 
\caption{\small{In general the energy of a network is infinite if the lower-dimensional D-brane tendrils are semi-infinite.  This infinity must be regularized, for example imposing an IR cutoff in the background, which may be automatic if the tendrils end on a horizon or the end of the world.  On the left we see a network which ends on a background brane but has infinite energy because the D$p$-brane's Dirac monopole is localized on a codimension 3 surface.  On the right a finite-energy BIon solution is drawn, which is in the same $\hat\partial$-homology class as that on the left, but the D$p$-D$(p-2)$ system has been replaced by a single continuous D$p$ that ends on the same background brane.}}
\label{bion}
\end{figure}

In the singular configurations that we have been discussing, in which a D$p$-brane ends on a D$(p+2)$-brane, the Dirac-Born-Infeld energy density of the D$(p+2)$-brane diverges at the points where it meets the D$p$-brane, giving an  infinite contribution to the energy.  This reflects the fact that these configurations contain a Dirac monopole in the D$(p+2)$-brane worldvolume theory, and the Dirac monopole solution has an infinite energy. Of course, the validity of the low-energy description provided by the DBI action breaks down and we should consider derivative corrections which are higher order  in $\alpha^\prime$.  However, these infinite-energy configurations are in the same $\hat{\partial}$-homology class as BIon-like configurations, which in flat space are smooth, all-order string theory solutions consisting of a single D$(p+2)$-brane tube \cite{thorlacius}.  Thus we recover the fact that the singular brane junctions relax to smooth, BIon solutions.  Both configurations are depicted in Fig.~\ref{bion}. In particular, this indirectly implies that the `topological' quantity
\bea\label{bigtheta}
\Theta_{[(\frakc,\frakf)]} \equiv \int_X\langle \omega,j_{(\frakc,\frakf)}\rangle\leq E(\frakc,\frakf)\ ,
\eea
which equals the energy of the supersymmetric configuration (if there is any) in the $\hat\partial$-homology class of $(\frakc,\frakf)$, is finite (when IR regularized) even if $\calf_\calc$ diverges on $\calc$, in agreement with the more general argument given after (\ref{PD}).

As a simple example, let us check our formalism by describing the  instability of $n$ D$0$-branes inside of a D$2$-brane wrapping a 2-cycle in a Calabi-Yau manifold.  It is known that it is energetically favorable for the D$0$-branes to dissolve in the D$2$-brane.  In this case, the appropriate \generalized\ calibration \cite{paulk,luca1} is given by
\bea
\omega=\Re(e^{i\theta}e^{-iJ})\ ,
\eea
where $J$ is the K\"ahler form of the Calabi-Yau and $\theta$ is a real parameter. 
We start with a D0-D2 configuration wrapping the generalized cycle $n(p_0,0)+(\Sigma,0)$, where $p_0$ is a point and $\Sigma$ a 2-cycle in the Calabi-Yau. The \generalized\ chain relating this configuration to the generalized cycle $(\Sigma,\calf)$, where $\int_\Sigma\calf=n$, was described at the end of section \ref{genhom}. Thus we can check that both $n(p_0,0)+(\Sigma,0)$ and $(\Sigma,\calf)$, when inserted into (\ref{bigtheta}), yield the same 
\bea
\Theta(\theta)=n\cos\theta+\sin\theta\int_\Sigma J\ 
\eea
where $J$ is the pullback of the K\"ahler form onto $\Sigma$.  

Note that in this case  we have a family of calibrations parametrized by the parameter $\theta$. In general, the correct $\theta$ gives the maximum value of $\Theta(\theta)$ and is selected by the condition
\bea
\int_X\langle \Im(e^{i\theta}e^{-iJ}),j_{(\frakc,\frakf)}\rangle=0. \label{calcond}
\eea
Such selection conditions are typically associated to backgrounds with extended supersymmetry that can be broken in different ways (like in our Calabi-Yau example) and  can be interpreted as imposing the vanishing of a Fayet-Iliopoulos term (for a discussion that includes nontrivial fluxes, see \cite{luca2}). 

In our example, (\ref{calcond}) is solved by
\bea
\theta=\tan^{-1}\frac{\int_\Sigma J}{n}.
\eea
Since $J$ is a standard calibration for two-cycles,  the energy associated to the first configuration $E=n+\text{Vol}(\Sigma)$ is always strictly bigger then $\Theta(\theta)$.  On the other hand, the second configuration can be calibrated if $\Sigma$ is holomorphically embedded and 
\bea
\sin\theta\,\calf=\cos\theta\, J\ .
\eea 
Thus, the first configuration will eventually continuously decay to the second and the D0-branes dissolve.

As a second check of our formalism, that also clarifies some previous comments about the energetics of D-branes ending on D-branes, let us consider the BIon solution in flat space, describing $n$ D1-branes ending on a D3-brane. The (ordinary) calibration for a D3-brane extending in the directions $x^1,x^2,x^3$ is given by $\omega_{\rm D3}=\d x^1\wedge \d x^2\wedge \d x^3$, while the calibration for the D1-branes extending along the $x^4$ direction is given by $\omega_{\rm D1}=\d x^4$. Thus the generalized calibration is given by
\bea
\omega=\omega_{\rm D3}+\omega_{\rm D1}=\d x^1\wedge \d x^2\wedge \d x^3 +\d x^4\ .
\eea
If $\sigma^\alpha$, $\alpha=1,2,3$, denote the three worldvolume coordinates on the D3-brane, we can choose a static gauge in which $x^\alpha=\sigma^\alpha$ and $x^4=X(\sigma)$. In general we allow a magnetic field strength $\calf=\frac12 \epsilon_{\alpha\beta\gamma} b_\gamma\d\sigma^\alpha\wedge\d\sigma^\beta$. Suppose that the flat space is IR regularized so that we have a finite volume $V_{\rm D3}$ along the three directions $x^1,\ x^2,\ $and $x^3$, while the fourth direction is bounded $|x^4|\leq L$. Then, if  the D3-brane  is located at $x^4=0$, $\Theta$ as defined in (\ref{bigtheta}) takes the finite value
\bea\label{thetaBI}
\Theta=V_{\rm D3}+nL\ .
\eea
This is indeed the energy of the D-branes, if we ignore the field strength on the $D3$-brane and should be identified with the energy of the system once it is relaxed to its supersymmetric configuration. The key point is that RR charge conservation demands a nontrivial field strength $\calf$, since the D1-branes act as monopoles and thus if, for example, the D3 and the D1-branes meet at $x^1=x^2=x^3=0$,  we must have  $\d\calf=n\delta^3(0)$. The configuration with straight D-branes cannot be supersymmetric, since the Dirac monopoles' energy diverges and indeed we know that a BIon should form. 
 
Let us recover the finite energy configuration by requiring the saturation of the calibration condition (\ref{calineq}) for the D3-brane. Since
\bea
[\omega|_{\rm{BIon}}\wedge e^\calf]_{\rm top}&=&\d x^1\wedge \d x^2\wedge \d x^3|_{\rm{BIon}}+\calf\wedge \d x^4|_{\rm{BIon}}=\cr &=&(1+\vec b\cdot\vec\nabla X)\,\d\sigma^1\wedge\d\sigma^2\wedge \d\sigma^3\ .
\eea    
and 
\bea
\cale=\sqrt{1+\vec\nabla X\cdot \vec\nabla X+\vec b\cdot\vec b +(\vec b\cdot \vec\nabla X)^2}\,\d\sigma^1\wedge\d\sigma^2\wedge \d\sigma^3\ ,
\eea
it is easy to see that the bound in (\ref{calineq}) is saturated only if $\vec\nabla X=\vec b$. Thus $\triangle X=0$ and, from the additional condition  $\d\calf=\vec\nabla \cdot\vec b\, \d^3\sigma= n\,\delta^3(0)$, we get $X=-n/(4\pi \rho)$, where $\rho^2=\vec\sigma\cdot\vec\sigma$. This is indeed the BIon solution, where the D3-brane extends to infinity, practically reabsorbing the D1-brane. This configuration is  in the same $\hat\partial$-homology class as the configurations with straight branes, and indeed its energy is given by (\ref{thetaBI}).

\section{The case of $\caln=1$ flux compactifications} \label{fluxsec}

We would now like to apply the previous arguments to $\caln=1$ flux compactifications to (warped) flat four-dimensional spacetime. We will focus on the backgrounds called D-calibrated in \cite{luca2}, since several general properties of D-branes on these kinds of flux backgrounds are  well understood \cite{luca1,luca2,lucadef}. 
In this case  $\calm=\mathbb{R}^3\times M$ and all background fields  preserve the four-dimensional Poincar\'e invariance. Letting $x^\mu=(t,x^i)$ and $y^m$  parametrize $\mathbb{R}^{1,3}$ and  $M$ respectively, the ten-dimensional metric can be written
\bea
\d s_{(10)}^2=e^{2A(y)}\d x^\mu\d x_\mu +g_{mn}(y)\d y^m\d y^n\ ,
\eea  
where $A$ is a possibly nontrivial warp factor depending only on the internal coordinates and $g_{mn}$ is the Euclidean metric on $M$.

Again we collect all of the RR field strengths into a single polyform of definite parity $F=\d_HC$. We split these components into electric and magnetic parts $\tilde F$ and $\hat F$ respectively
\bea
F=\text{vol}_{4}\wedge\tilde F+\hat F\ ,
\eea
where $\text{vol}_{4}=e^{4A}\d t\wedge \d x^1\wedge \d x^2\wedge \d x^3$.
Note that $\tilde F$ and $\hat F$ are not independent  due to the self-duality relations between the RR field strengths. 
Therefore all information about the RR fluxes is contained in either $\tilde F$ or $\hat F$, depending on whether we choose the electric or magnetic description.  If we treat $\tilde F$ as the fundamental field then we can locally introduce  `electric' RR gauge fields, and choose a gauge in which they take the form 
$C=\text{vol}_{4}\wedge \tilde C$, with $\tilde F=e^{-4A}\d_H (e^{4A} \tilde C)$. If $\hat F$ are the fundamental field strengths, the corresponding `magnetic' RR gauge fields are $C=\hat C$, so that $\hat F=\d_H \hat C$.

The supersymmetry conditions for these backgrounds were studied in \cite{gmpt}, where it was shown that they are characterized by two complex polyforms $\hat\Psi_1$ and $\hat\Psi_2$ which, using the terminology of generalized complex geometry \cite{hitchin,gualtieri}, are compatible pure-spinors  (they square to the volume  form on $M$ and define an $SU(3)\times SU(3)$ structure group on $TM\oplus T^\star\!M$).\footnote{We use the normalization for the pure spinors introduced in \cite{luca2}.} The background supersymmetry conditions  can be written
\bea\label{backsusy}
\d_H(e^{4A-\Phi}\Re \hat\Psi_1)=e^{4A}\tilde F\quad,\quad \d_H(e^{2A-\Phi}\Im \hat\Psi_1)=0\quad,\quad \d_H(e^{3A-\Phi}\hat \Psi_2)=0\ .
\eea
As was shown in \cite{luca1}, and further discussed in \cite{luca2}, the three conditions (\ref{backsusy}) are associated to the calibrations for D-branes wrapping an internal generalized cycle $(\Gamma,\calf)$ of definite dimension and filling respectively four, two and three flat dimensions. These three generalized calibrations are summarized by a single generalized calibration defined on $\calm$
\bea\label{totcal}
\hat\omega &=&\d x^{1}\wedge \d x^2\wedge \d x^3\wedge e^{4A-\Phi}\Re \hat\Psi_1+ \hat n\wedge e^{2A-\Phi}\Im \hat\Psi_1+\cr &&+m^{(1)}\wedge e^{3A-\Phi}\Re(e^{i\theta}\hat\Psi_2) +m^{(2)}\wedge e^{3A-\Phi}\Im(e^{i\theta}\hat\Psi_2)\ ,
\eea
where $\hat n=\hat n_i\d x^i$, $ m^{(1)}=\frac12 m^{(1)}_{ij}\d x^i\wedge \d x^j$ and  $m^{(2)}=\frac12 m^{(2)}_{ij}\d x^i\wedge \d x^j$ are constant forms that are constrained as follows. Define the constant one-forms   $\hat m^{(1)}=\frac12 \epsilon_{ijk}m^{(1)}_{ij}\d x^k$ and $\hat m^{(2)}=\frac12 \epsilon_{ijk}m^{(2)}_{ij}\d x^k$. Then  we must impose the normalization conditions $\hat n\cdot \hat n=\hat m^{(1)}\cdot \hat m^{(1)}=\hat m^{(2)}\cdot \hat m^{(2)}=1$, plus the compatibility conditions $\hat m^{(1)}\cdot \hat m^{(2)}=0$ and $\hat n= \hat m^{(2)}\times \hat m^{(1)}$. Using (\ref{backsusy}) one can see that the condition (\ref{diffcond2}) is satisfied. The calibration (\ref{totcal}) contains four independent parameters, $\theta$ plus the three parameters contained in $m^{(1)}$ and $m^{(2)}$, which are fixed by the particular homology class  of the generalized cycles  one wants to calibrate. The one form $\hat n$ identifies the direction along which a D-string extends, while $\hat m^{(1)}$ and $\hat m^{(2)}$ are orthogonal to the directions that can be spanned by possible domain walls.\footnote{A general calibrated network of domain walls and strings preserves $1/4$ of the underlying four-dimensional $\caln=1$ supersymmetry, defined by the conditions $e^{i\theta\gamma_5}(\hat m^{(1)}\cdot \vec \gamma)\xi=\xi$ and $(\hat n\cdot \vec\gamma)\gamma_0\xi=\xi$, where $\xi$ is the surviving background four-dimensional Majorana Killing spinor.} 

The calibration (\ref{totcal}) is the most general, but one can always use Poincar\'e invariance to fix three of the parameters, e.g. we can set $\hat n= \d x^1$, $m^{(1)}=\d x^1\wedge \d x^2$ and $m^{(2)}=\d x^3\wedge\d x^1$. Also, by a chiral rotation, one can fix the angle $\theta=0$.  With this choice of coordinates, all strings and domain walls must extend along the $x^1$ direction and the domain walls also extend along another direction in the $23$-plane. The relevant calibration for these domain walls is
\bea\label{redcal}
\d x^1\wedge \d x^2\wedge e^{3A-\Phi}\Re\hat\Psi_2 +\d x^3\wedge \d x^1\wedge e^{3A-\Phi}\Im\hat\Psi_2\ .
\eea 
Here we have fixed the phase $\theta$ so that the internal generalized cycle  is calibrated by $e^{3A-\Phi}\Re\hat\Psi_2$ or $e^{3A-\Phi}\Im\hat\Psi_2$  if the domain wall fills the $12$- or $31$-directions respectively. However, we can also consider   domain wall embeddings of the form
\bea
x^1=\sigma^1\quad,\quad x^2=\sigma^2\cos\alpha\quad,\quad x^3=\sigma^2\sin\alpha\ ,
\eea
where $\sigma^1$\ and $\sigma^2$ are flat worldvolume coordinates and $\alpha$ is a constant angle. From (\ref{totcal}) one can easily see that in this case the domain wall must wrap a generalized cycle calibrated by $e^{3A-\Phi}\Re(e^{i\alpha}\hat\Psi_2)$. Thus a rotation around the $1$-axis  must be accompanied by a chiral rotation of the preserved Killing spinor, in agreement with the purely algebraic arguments of \cite{tow}.

We may consider networks of D-branes, where for example, from the four-dimensional point of view, a space-filling brane ends on a domain wall, or a domain wall ends on a string, or two or more non-parallel domain walls intersect or end on the same string, etc.  The possible supersymmetric D-brane networks of this kind depend on the details of the background considered. In the next section we will analyze the case of the type IIB  $\caln=1$ compactifications on warped Calabi-Yau's with fluxes as discussed for example in~\cite{gkp}. 

\section{Domain wall networks on warped Calabi-Yaus} \label{cysec}

In general, in a type II $\caln=1$ flux background  the internal manifold is not necessary a Calabi-Yau and may not even be complex or symplectic, but the supersymmetry implies that the internal space is a generalized Calabi-Yau \`a la Hitchin \cite{hitchin,gmpt}. However, there is a class of IIB flux backgrounds which preserve the internal Calabi-Yau, where the net effect of the fluxes results only in a warping of the flat and internal metric \cite{gkp,gp}.  The axion-dilaton $\tau=C_{(0)}+ie^{-\Phi}$ is constant (with $e^\Phi=g_s$) and $G_{(3)}=\hat F_{(3)}+\tau H$ is $(2,1)$ and primitive, with $\hat F_{(3)}$    quantized analogously to the $H$ flux (see Eq.~(\ref{quantcond}) and appendix \ref{RRquant}). The localized D3-brane charge density $\rho_{\rm D3}$ must obey the tadpole condition
\bea
\d \hat F_{(5)}= \hat F_{(3)}\wedge H+\rho_{\rm D3}^{\rm loc}\ , 
\eea 
which in particular implies that $Q_{\rm D3}=\int_M H\wedge \hat F_{(3)}$.
In this case we have\footnote{In this paper the K\"ahler form $J$ differs by an overall sign from the one  used in \cite{luca1,luca2}.}
\bea
\hat\Psi_1=e^{iJ}\quad,\quad \hat\Psi_2=\Omega
\eea 
where $J$ and $\Omega$ are a $(1,1)$ and $(3,0)$ form on the internal manifold, such that 
\bea
J\wedge\Omega=0\quad,\quad \frac1{3!}J\wedge J\wedge J=-\frac i8 \Omega\wedge\bar\Omega\ . 
\eea
The background supersymmetry conditions (\ref{backsusy}) imply that $\hat J\equiv e^{2A}J$ and $\hat \Omega\equiv e^{3A}\Omega$ are closed and define the K\"ahler and the holomorphic $(3,0)$-form associated to the internal Calabi-Yau manifold.  In this case, fixing the chiral symmetry and Poincar\'e invariance as in the previous section, the total calibration is given by\footnote{To simplify notation,  we do not write explicitly the overall constant factor $e^{-\Phi}=1/g_s$ in the following expressions involving calibrations.} 

\bea\label{wcycal}
\hat\omega_{\rm WCY}&=&\d x^1\wedge \d x^2\wedge \d x^3\wedge (e^{4A}-\frac12\hat J\wedge \hat J)+\d x^1\wedge (\hat J - \frac{1}{3!}e^{-4A}\hat J\wedge \hat J\wedge \hat J)+\cr 
&&+\d x^1\wedge \d x^2\wedge \Re \hat\Omega+\d x^3\wedge \d x^1\wedge \Im \hat\Omega\ .
\eea

Consider first a single domain wall obtained by wrapping a D5-brane on an internal 3-cycle $\Gamma$. Having fixed the Poincar\'e invariance and chiral symmetry, a BPS domain wall must fill the $x^1$ direction and form an angle $\alpha(\Gamma)$ with the $x^2$ axis in the $(x^2,x^3)$-plane given by\footnote{Notice that, in contrast with the standard assumption, $\hat \Omega$ has both fixed normalization and  fixed phase due to the condition on the chiral phase $\theta$.}
\bea\label{angledw}
\alpha(\Gamma)=\arg \Big(-\int_\Gamma\hat\Omega\Big)\ .
\eea   
The tension of the domain wall is then given by $T_{\rm DW}(\Gamma)=|\int_\Gamma\hat\Omega|$. From (\ref{modBI}), the worldvolume tadpole condition implies that we must add  $N_{\rm D3}(\Gamma)=\int_{\Gamma}H$ D3-branes filling half of spacetime and ending on the domain wall. The D5-brane creates a jump in the $\hat F_{(3)}$ flux that exactly compensates for the variation of the number of D3-branes so that  $\Delta Q_{\rm D3}=\int_M H\wedge \Delta \hat F_{(3)}$ and so the background tadpole condition is satisfied on both sides of the domain wall. We can also consider configurations obtained by superimposing different domain walls at different angles. 

It is worthwhile to make an observation at this point. When the internal space $M$ is compact, one may expect an effective four-dimensional $\caln=1$  supergravity description of the above D-brane domain walls. In four-dimensional  supergravity, domain walls interpolating between two Minkowski vacua have vanishing tension \cite{cvetic,ferrara,kallosh}. Although the existence of such a  four-dimensional description of D-brane domain walls is not guaranteed (for a discussion see \cite{dealwis1,dealwis2}),  the same kind of problem arises in our formalism. We have seen that the tension of a supersymmetric domain wall obtained by wrapping a D5-brane around an internal cycle $\Gamma$ is given by
\bea
T_{\rm DW}(\Gamma)=\frac{1}{g_s}|\int_\Gamma\hat\Omega|.
\eea
Since the domain wall induces a jump  in the RR-flux that in  cohomology is given by $[\Delta\hat F_{(3)}]={\rm PD}_M([\Gamma])$, if $\hat\Omega$ is constant as one traverses the wall, one finds
\bea
T_{\rm DW}=\frac{1}{g_s}|\int_M\hat\Omega\wedge \Delta\hat F_{(3)}|.
\eea
 This means that in order to have $T_{\rm DW}\neq 0$ one must necessarily have  $ \Delta\hat F_{(3)}^{(0,3)}=\Delta\hat F_{(3)}^{(3,0)}\neq 0$ and then the jump in the RR flux generated by the domain wall seems to break the supersymmetry of the background.    On the other hand, if one considers the calibrated D-branes as probes that do not deform the background, they exactly preserve part of the background supersymmetry by standard $\kappa$-symmetry arguments. Thus,  when the probe approximation for the D5-brane applies, one may consider the supersymmetry breaking effect induced by the flux jump as subleading and  the resulting domain wall configurations as quasi-supersymmetric. Domain-walls from 5-branes  can have several applications, as  discussed for example in \cite{kachru2,kpv,verlinde,dealwis1,dealwis2}.\footnote{See also \cite{marchesano} for a discussion of interesting physical effects due to the twisting induced by the $H$-flux in IIB flux compactifications and their IIA mirrors.}

More interesting configurations may be obtained by gluing together several BPS domain walls that extend at different angles along a half-plane and meet along a line. 
Suppose first that we want to directly glue together a number of D5-branes wrapping different cycles $\Gamma^{(i)}$. The necessary topological condition is that $\sum_i\Gamma^{(i)}$ is trivial in homology. As a result $\sum_i\int_{\Gamma^{(i)}}\hat\Omega=0$ and so
\bea\label{anglecond}
\sum_i T_{\rm DW}(\Gamma^{(i)})\cos\alpha(\Gamma^{(i)})=0\quad,\quad \sum_i T_{\rm DW}(\Gamma^{(i)})\sin\alpha(\Gamma^{(i)})=0\ .
\eea
These equations have a simple interpretation as the equilibrium conditions that the tension of each domain wall must be compensated  by the other domain walls.  
Regarding the numbers $N_{\rm D3}(\Gamma^{(i)})$ of D3-branes that can be inserted, the triviality of $\sum_i\Gamma^{(i)}$ also implies that 
\bea\label{d3cond}
\sum_i N_{\rm D3}(\Gamma^{(i)})=0\ .
\eea
This constraint also has a clear geometrical interpretation. A semi-infinite domain wall wrapping a cycle $\Gamma^{(i)}$ which supports a nontrivial $H$-flux must have $N_{\rm D3}(\Gamma^{(i)})$ more spacetime filling D3-branes on one side than the other. The condition (\ref{d3cond}) simply requires that the $H$-flux on the other cycles is such that this condition can be simultaneously satisfied on all walls. In other words, if for example we have a junction of three semi-infinite domain walls, crossing the first wall, then the second, then the third, one returns to the original region and so the total jump in D3 brane charge must vanish.  Finally, let us also note that the triviality of  $\sum_i\Gamma^{(i)}$ obviously implies that the total jump of the $\hat F_{(3)}$ after a complete path around the junction is trivial since $[\Delta^{\rm tot}\hat F_{(3)}]=\sum_i\text{PD}_M[\Gamma^{(i)}]=0$.


Consider now a string extending along the $x^1$-direction obtained by wrapping a D7-brane on the internal manifold. By itself such a configuration cannot be consistent since the $H$-flux on the worldvolume of the D7-brane is nontrivial. However, we can attach to the string several domain walls that can be at different angles and are obtained by wrapping D5-branes on  three-cycles $\Gamma^{(i)}$. The field strength $\calf$ on the internal part of the D7-brane must then satisfy the modified Bianchi identity $\d\calf=H+\sum_i\delta^3_M(\Gamma^{(i)})$ and thus the consistency of the configuration requires that, in homology, $\sum_i \text{PD}_M([\Gamma^{(i)}])=-[H]$. This in turn implies again (\ref{d3cond}) with the subsequent interpretation as a topological condition on the spacetime filling D3-branes. Note also that the background supersymmetry conditions require that $\hat\Omega\wedge H=0$. It follows that
\bea\label{Hangle}
\sum_i\int_{\Gamma^{(i)}}\hat\Omega=\int_M\hat\Omega\wedge H=0\ ,
\eea
and we obtain again  the equilibrium conditions (\ref{anglecond}).  

Notice that in this case the total jump in  $\hat F_{(3)}$ as one crosses the domain walls does not vanish in cohomology, since
\bea
[\Delta_{\rm DW's}^{\rm tot}\hat F_{(3)}]=\sum_i\text{PD}_M([\Gamma^{(i)}])=-[H].
\eea
However this does not mean that such a configuration is inconsistent. Indeed, the monodromy of $C_{(0)}$ as one encircles the D7-brane is equal to one unit. This induces a variation
\bea
[\Delta^{\rm tot}_{\rm string}\hat F_{(3)}]=(\Delta^{\rm tot}C_{(0)})[H]=[H]
\eea
that  exactly corrects the total jump created by the domain walls. Thus the total variation of $\hat F_{(3)}$ as one encircles the D7 is trivial, so $\hat F_{(3)}$ is globally defined and the configuration is consistent.  Such a cancellation in the variation of improved field strengths always occurs~\cite{memono}.


\subsection{An example: networks on  $T^6/\mathbb{Z}_2$ flux compactifications}
\label{Torient}

As a simple example, let us focus on  the $T^6/\mathbb{Z}_2$ orientifold flux compactifications studied in \cite{kachru}.  Start with type I string theory on $\mathbb{R}^{1,3}\times T^6$ where $T^6$ is the six-torus and T-dualize all six circles.  The result is type IIB on $T^6/\Z_2$ with 64 O3-planes.\footnote{As described in \cite{HK}, there are four varieties of O3 planes, characterized by a possible NSNS and a possible RR $\Z_2$-valued discrete torsion on the linking $\mathbf{RP}^5$.  We will consider a background in which both discrete torsions on all 64 planes are turned off.}  Such planes have tensions equal to minus one quarter of the D3-brane tension. The internal three-form fluxes and  the number $N_{\rm D3}$ of  D3-branes which extend along the noncompact $\mathbb{R}^{1,3}$ are related  by the tadpole condition
\bea
N_{\rm D3}=16+\frac{1}{2}\int_{T^6}H\wedge \hat F_{(3)} \label{tad}\ ,
\eea
where $16=64/4$ is the contribution of the 64 orientifold planes and the factor of one half reflects the fact that one need integrate over only one fundamental domain $T^6/\Z_2$ of the $\Z_2$ action on $T^6$.

In \cite{kachru}, a class of supersymmetric vacua was discussed, in which  the fluxes $H$ and $\hat F_{(3)}$ have constant components in the 6-torus coordinates $x^a\simeq x^a+1$ and $y^a\simeq y^a+1$, the complex structure is defined by the complex coordinates $z^a=x^a+\phi y^a$ and the associated (non-normalized) holomorphic $(3,0)$-form is  $\hat\Omega= \d z^1\wedge \d z^2\wedge \d z^3$. A family of compatible K\"ahler structures is given by $\hat J= -i\sum_a r_a^2\,\d z^a\wedge \d \bar z^a\sim \sum_a r_a^2\,\d x^a\wedge \d y^a$, for arbitrary $r_a$'s. We now want to consider a BPS probe D5-brane wrapped on a three cycle $\Gamma$ on the $T^6/\mathbb{Z}_2$, that for simplicity we assume descends from a cycle (and its image under the orientifold involution) on $T^6$. Thus $\Gamma$ must be calibrated by $\Re (e^{i\theta}\hat\Omega)$ for some $\theta$, i.e. it must be a special Lagrangian (SLag) cycle, as will be discussed in general in section \ref{2exsec}. This  implies that the pullback of $\hat J$ to $\Gamma$ must vanish (i.e. $\Gamma$ is a Lagrangian cycle) and $\Im(e^{i\theta}\hat\Omega|_\Gamma)=0$ (the `special' condition). Supersymmetric cycles are thus described by an embedding of the form $\sigma^a\mapsto (x^a=m^a\sigma^a+c^a,y^a=n^a\sigma^a+d^a)$ (no sum over repeated indices), where the $\sigma^a$'s denote the worldvolume coordinates and $m^a,n^b\in \mathbb{Z}$ are winding numbers. We will denote such supersymmetric cycles by $\Gamma_{\vec m,\vec n}$, where a choice of $c^a$ and $d^a$ is implicit.  The phase $e^{i\theta}$ is fixed by the condition $\Im(e^{i\theta}\hat\Omega|_\Gamma)=0$ up to a sign and there must be $N^{\vec m,\vec n}_{\rm D3}=|\int_{\Gamma_{\vec m,\vec n}} H|$ D3-branes ending on a D5 that wraps $\Gamma_{\vec m,\vec n}$.

\begin{figure}[ht]
\begin{center}
\leavevmode
\epsfxsize 10   cm
\epsffile{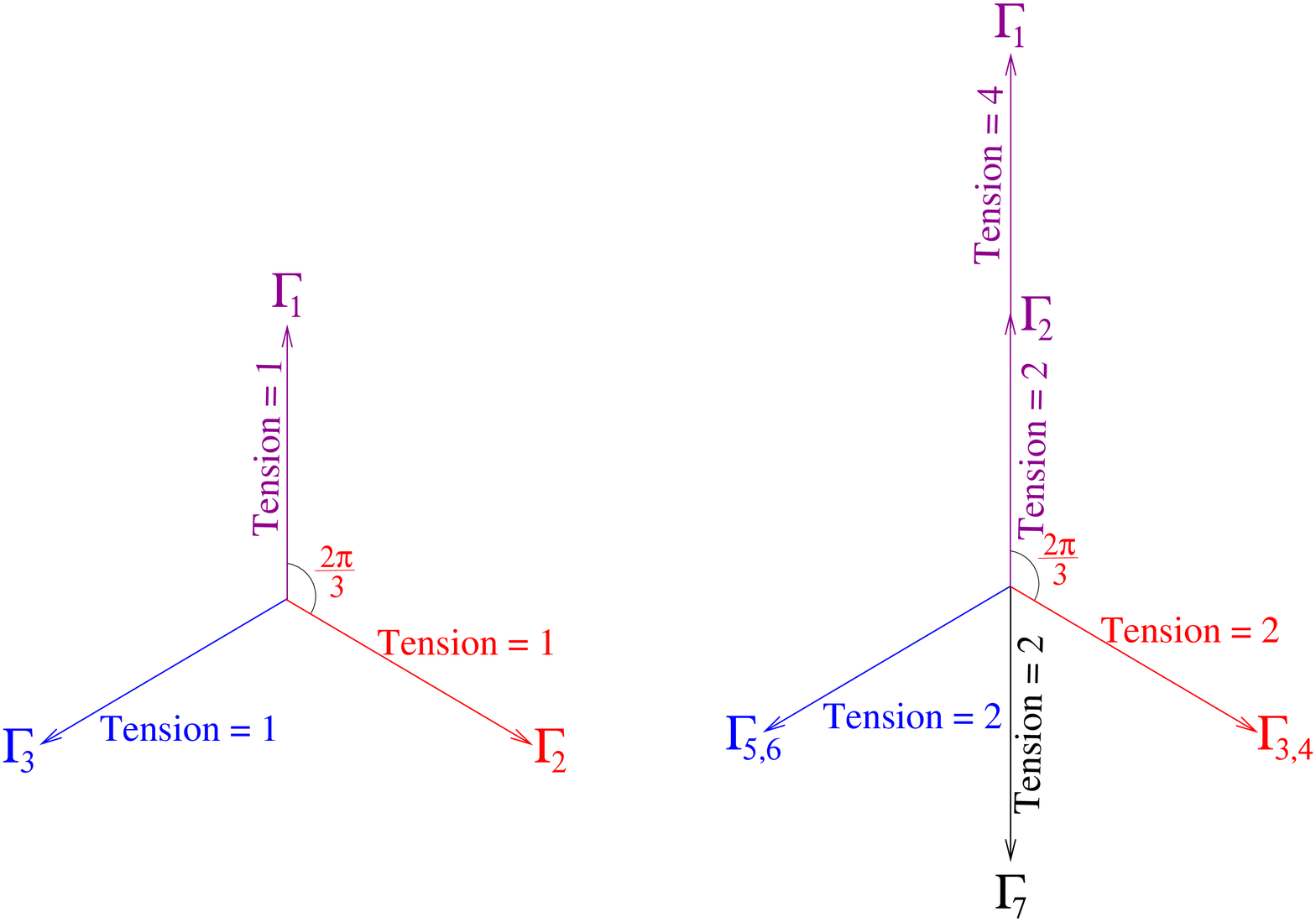}    
\end{center} 
\caption{\small{The $(x^2,x^3)$ cross-section of two examples of networks of domain walls are drawn, indicating explicitly the contribution to the total wall tensions coming from  each D5-brane in units where $\text{Vol}(T^6)/g_s^2=1$.  On the left there are three walls of tension equal to one in the four-dimensional theory, which in the full theory are D5-branes that wrap the cycles $\Gamma_1=A^0$, $\Gamma_2=A^1$ and a third cycle $\Gamma_3$ which is homologous to $-A^0-A^1$.  These extend at relative angles of 120 degrees in the $(x^2,x^3)$-plane, and so their tensions cancel in the four-dimensional sense leaving a stable configuration, in agreement with the general formula (\ref{anglecond}).  On the right 7 intersecting domain walls are drawn, which extend in four distinct directions in this 2-dimensional cross-section, although they all wrap distinct cycles on the internal toroidal orientifold.  These walls all end on a string which lifts to a D7-brane.  The fact that $\hat{\Omega}\wedge H$ integrates to zero, combined with the fact that the total cycle wrapped by the walls is equal to $-H$ according to the Freed-Witten anomaly, guarantees that the tensions of these walls also cancel (see Eq.~(\ref{Hangle}) and related discussion), as can be verified explicitly in the figure.}}
\label{murifig} 
\end{figure}

We will  focus on the following explicit background, described in \cite{kachru}. Let us first introduce the following real integer three-forms on $T^6$
\bea
&&\alpha_0=\d x^1\wedge\d x^2\wedge\d x^3\quad,\quad \beta^0=\d y^1\wedge\d y^2\wedge\d y^3\ ,\\
&&\alpha_a=\frac12\sum_{b,c} \epsilon_{abc}\d y^a\wedge\d x^b\wedge\d x^c\quad,\quad\beta^a=-\frac12 \sum_{b,c}\epsilon_{abc}\d x^a\wedge\d y^b\wedge\d y^c\nonumber\ ,
\eea 
where, in the last line, no sum over the indices $a$ is taken.  In terms of these forms the background of interest supports the following  three-form fluxes
\bea
\hat F_{(3)}&=&2\alpha_0+2\sum_{a=1}^3\alpha_a+2\sum_{a=1}^3\beta^a+8\beta^0\ ,\cr
H&=&-2\alpha_0+2\sum_{a=1}^3\alpha_a+2\sum_{a=1}^3\beta^a+4\beta^0\ .
\eea
The axion-dilaton is $\tau=2e^{2\pi i/3}$ while the complex structure is given by $\phi=e^{2\pi i/3}$, and thus\footnote{Note that  in our and \cite{kachru}'s notation, $\tau$ and $\phi$ have opposite meaning.}
\bea\label{toruscs}
\hat\Omega=\alpha_0+e^{\frac{2\pi i}{3}}\sum_{a+1}^3\alpha_a+e^{\frac{i\pi }{3}}\sum_{a=1}^3\beta^a+\beta^0\ .
\eea

We can choose eight SLag cycles $A^I, B_J$ of the kind $\Gamma_{\vec m,\vec n}$ discussed above, with $I,J=0,\ldots,3$, such that\footnote{For example $A^0$ corresponds to $m^1=m^2=m^3=1$ and $n^1=n^2=n^3=0$, $A^1$ to $n^1=m^2=m^3=1$ and $m^1=n^2=n^3=0$, and so on.} 
\bea
\int_{A^I} \alpha_J=\int_{B_J}\beta^I=\delta^I_J\ ,
\eea 
and construct domain wall networks by wrapping D5-branes around  $A^I$ and $B_J$. Consider first the case in which we do not need a D7-brane insertion at the junction. Thus, as we have discussed, the topological requirement is that the sum of the different cycles wrapped by the D5-branes is trivial in homology. A simple example of such a configuration is provided by a junction of three D5-branes wrapping the internal SLag cycles $\Gamma_1=A^0$, $\Gamma_2=A^1$ and $\Gamma_3$, which is of the form $\Gamma_{\vec m,\vec n}$ where $m^1=n^1=-m^2=-m^3=-1$ and $n^2=n^3=0$. Thus $[\Gamma_3]=-[\Gamma_1+\Gamma_2]$. Using (\ref{toruscs}) and (\ref{angledw}), one can easily see that the second and the third domain walls  are at angles $\alpha=2\pi/3$ and $-2\pi/3$ respectively on the $(x^2,x^3)$-plane with respect to the first domain wall. This configuration can be seen on the left of Fig.~\ref{murifig}.  This example can be straightforwardly extended to more complicated configurations involving more then three D5-brane walls intersection along a common string-like boundary.

We can also construct networks of the second kind discussed above, where several D5-branes wrapping internal cycles $\Gamma_i$ meet along a string-like junction filled by a D7-brane wrapping the entire internal space $M$.  In this case the topological condition $\sum_i \text{PD}_M([ \Gamma_i])=-[H]$ is satisfied by the following seven SLag cycles 
\bea
&&\Gamma_1=-4A^0\quad,\quad \Gamma_2=-2 B_0\quad,\quad \Gamma_{3}=- 2A^2\quad,\quad \Gamma_{4}=- 2A^3\ ,\cr &&\quad\quad\quad\quad\quad\Gamma_{5}=2B_1\quad,\quad \Gamma_{6}=2B_3\quad,\quad \Gamma_7\ ,
\eea
where $\Gamma_7$ is $\Gamma_{\vec m,\vec n}$ with $m^1=n^2=m^3=n^3=1$ and $n^1=m^2=0$.
From (\ref{toruscs}) and (\ref{angledw}) one can see that we have four stacks of D5-branes extending along four different directions on the $(x^2,x^3)$-plane. The first group consists of two D5 branes wrapping the cycles $\Gamma_1$ and $\Gamma_2$ and let us say that they are at angle $\alpha=0$. The second group is given by the two D5-branes wrapping the cycles $\Gamma_{3}$ and $\Gamma_4$, which are at an angle $2\pi/3$.  The third group is given by the two D5-brane wrapping the cycles $\Gamma_{5}$ and $\Gamma_6$, which are at an angle $-2\pi/3$. Finally we have the D5-brane wrapping the cycle $\Gamma_7$ at an angle $\pi$. This configuration is depicted on the right of Fig.~\ref{murifig}.

Two final observations are in order. First, we would like to stress that the above  D-brane networks are supersymmetric only in the probe approximation, where the background is kept fixed and the D-branes' backreaction is neglected. However, in the present  example the string coupling constant is $g_s=1/\sqrt{3}\simeq 0.58$ and the jump in the RR flux generated by the D5-branes is comparable with the background fluxes. Thus, while we expect this approximation to be justified in several interesting flux vacua, it cannot be trusted in our simple example and one should in principle face the aforementioned  problem concerning supersymmetry-breaking flux jumps generated by the domain walls. Second, we have not found the actual supersymmetric  D-brane network configuration, rather we have started from separated D3, D5 and D7-branes and we have checked that they can be glued together consistently, obtaining also information about tensions and angles from the calibration  (\ref{wcycal}). However, accordingly to the general discussion of section \ref{calisec}, the calibration  (\ref{wcycal}) can be also used to determine the actual BPS configuration inside the same $\hat\partial$-homology class  of each of these networks where lower-dimensional D-branes end on higher dimensional ones. In the next section, we will show that this is explicitly possible by concentrating on a single composite domain wall, where several D3-branes can end on a wall obtained by wrapping a D5-brane on an internal cycle supporting a non-trivial $H$-flux.


\section{The shape of composite domain walls} \label{2exsec}

Until now we have used generalized calibrations to extract information on D-brane networks of BPS domain walls and strings that essentially depends only on the topology of the configuration. However, imposing the calibration condition corresponding to  the saturation of the local bound (\ref{calineq}) (or equivalently (\ref{calineq2}))  should allow one to obtain the explicit D-brane configurations. We now show this, focusing on a domain wall corresponding to a D5-branes wrapping a three cycle $\Gamma$ supporting a possible nontrivial $H$-flux. 
Such domain walls can be for example relevant in the context of flux compactifications and the gauge/gravity correspondence (see e.g. \cite{kpv,kachru2,dealwis1,dealwis2,verlinde}).

If the $H$-flux on $\Gamma$ were trivial and we could set $\calf=0$, the calibration condition would imply that $\Gamma$ is a SLag cycle  \cite{luca1}. The presence of a nontrivial $H$-flux drastically changes the situation, since we must now attach $N_{\rm D3}(\Gamma)$ D3-branes filling half of spacetime  to the domain wall. They act as monopole sources on $\Gamma$ and thus $\calf$ cannot be trivial and in fact they deform the geometry of the D5-brane  by `pulling' it, creating BIon-like spikes.  

To analyze the resulting supersymmetric configuration more explicitly, it is convenient to use the warped metric $\hat g_{mn}(y)=e^{2A(y)}g_{mn}$ on the internal manifold $M$. The shape of the D5-brane is specified  by the embedding functions  $y^m=Y^m(\sigma)$, in the internal directions, and $x^3=X(\sigma)$, where $\sigma^\alpha$, $\alpha=1,2,3$, are coordinates on the three-cycle $\Gamma$ wrapped by the D5-brane. On $\Gamma$ we will use the pullback of the background internal warped metric $\hat g_{mn}$.  Using the calibration (\ref{wcycal}) with the phase $e^{i\theta}$ originally present in (\ref{totcal}) reintroduced and following a procedure almost identical to that  used for the BIon solution in section \ref{calisec}, the calibration condition $[\hat\omega|_\Gamma\wedge e^\calf]_{\text{top}}= \cale_{\text{DBI}}(\Gamma,\calf)$  implies that, for some $\theta$,
\bea\label{DWcond}
\Re(e^{i\theta}\hat\Omega)|_\Gamma=\sqrt{\det(\hat g|_\Gamma)}\,\d^3\sigma\ ,
\eea
plus the BPS condition relating the profile $X(\sigma)$ to the $\calf$ flux
\bea\label{bps}
\d X=\star_3\calf\ .
\eea
If the D3-branes end on $\Gamma$ at the points $p_i$, $i=1,\ldots,N_{\rm D3}(\Gamma)$, from the worldvolume Bianchi identities (\ref{modBI}) it follows that
\bea\label{lapldw}
\Delta X=\star_3[ H|_\Gamma+\sum_i\delta_\Gamma^3(p_i)]\ ,
\eea
where $\Delta=\star_3\d\star_3\d$ is the Laplacian operator. We thus see that the D5-brane extends in BIon-like spikes in the $x^3$ direction localized at the points $p_i$. 
Note however that (\ref{DWcond}) implies that the D5-brane still wraps a SLag internal three-cycle.

The general solution to equation (\ref{lapldw}) can be found in terms of an associated Green's function ${\cal G}(\sigma,\sigma^\prime)$, which must satisfy
\bea\label{green}
\Delta_\sigma{\cal G}(\sigma,\sigma^\prime)=\Delta_{\sigma^\prime}{\cal G}(\sigma,\sigma^\prime)=\star_3\delta_\Gamma^3(\sigma-\sigma^\prime)-\frac{1}{\ \hat{\text{V}}_3}\ ,
\eea
where $\hat{\text{V}}_3=\int_\Gamma\d^3\sigma\sqrt{\det(\hat g|_\Gamma)}=T_{\rm DW}(\Gamma)$. Thus
\bea\label{gensol}
X(\sigma)=X_0+\sum_i{\cal G}(\sigma,\sigma_i)+\int_\Gamma{\cal G}(\sigma,\sigma^\prime)H|_\Gamma(\sigma^\prime)\ ,
\eea
where $\sigma_i$ denotes the coordinates of the point where the $i$th D3-brane ends on the D5-brane and  $X_0$ is the arbitrary additive constant (the zero mode of the Laplacian operator) which corresponds to the position of the brane.

An important observation is that the final domain wall tension is still given by $|\int_\Gamma\hat\Omega|$, as in the case of trivial $H$-flux or no flux at all. This value coincides with the minimal volume that a cycle homologous to $\Gamma$ can have.  Indeed, the tension can be easily computed from the complete generalized calibration $\omega=\hat\omega-e^{4A}\tilde C$. The background supersymmetry conditions imply that we can choose $\tilde C_{(0)}=1$ and the relevant part of $\omega$ is given by
\bea
\omega=\d x^1\wedge \d x^2\wedge \Re\hat\Omega\ .
\eea   
Thus the domain wall tension does not depend at all on the nontrivial $X(\sigma)$ and $\calf(\sigma)$ but only on the internal embedding described by $Y^m(\sigma)$.
Note that in general we have a non-trivial $\calf$ flux, related to the $X(\sigma)$ profile by the BPS equation (\ref{bps}). This is of course unavoidable if there are D3-brane insertions, but also when $\int_\Gamma H=0$ with nevertheless non-vanishing $H|_\Gamma$. If we want a domain wall at a fixed position we must impose $\calf=0$ (and thus also $H|_\Gamma=0$), consistently with the results of \cite{luca1}. 

\subsection{Examples}
As a first example, let us take  the $T^6/\mathbb{Z}_2$ orientifold flux compactifications considered in subsection \ref{Torient}. We can solve equation (\ref{lapldw}), using the general formula (\ref{gensol}). The relevant  metric induced on an internal cycle $\Gamma_{\vec m,\vec n}$ is $\hat g|_{\Gamma_{\vec m,\vec n}}=\sum_a R^2_a\d \sigma^a\d \sigma^a$, where $R_a^2=|m_a+\phi n_a|^2 r_a^2$, so that $\hat V=R_1R_2R_3$. Thus the Green's function is given by
\bea
\calg(\sigma,\sigma^\prime)_{\Gamma_{\vec m,\vec n}}=-\frac{1}{\hat V}\sum_{\vec k\neq \vec 0}\frac{e^{2\pi i \vec k \cdot(\vec\sigma-\vec\sigma^\prime)}}{(2\pi)^2\sum_a(k_a/R_a)^2}\quad,\quad \vec k=(k_1,k_2,k_3)\in \mathbb{Z}^3\ .
\eea
Since the $H$-flux is constant in the $(x^a,y^a)$ coordinates, the second term on the right hand side of  (\ref{gensol}) vanishes and so we have the general solution for the shape of the domain wall
\bea\label{gentorus}
X(\sigma)=X_0-\frac{1}{\hat V}\sum_{\vec k\neq \vec 0}\sum_{i=1}^{N^{\vec m,\vec n}_{\rm D3}}\frac{e^{2\pi i \vec k \cdot(\vec\sigma-\vec\sigma_{i})}}{(2\pi)^2\sum_a(k_a/R_a)^2}\quad,\quad \vec k=(k_1,k_2,k_3)\in \mathbb{Z}^3\ .
\eea
If we restrict our attention to the isotropic case in which $R_1=R_2=R_3=R$, it is easy to see that, if we zoom the solution around some D3-brane insertion $\sigma_i$, we recover the BIon solution discussed in section \ref{calisec}, i.e. 
$X(\sigma)\sim X_0 +1 /(4\pi R|\vec\sigma-\vec\sigma_i|)$ for $|\sigma-\sigma_i|<< 1$,  as expected. 

As second example, we consider a case where  the internal space is non-compact.\footnote{Thus, in this case, the mentioned subtleties in the probe approximation and the possible supersymmetry-breaking backreaction do not arise.} Type IIB flux vacua involving an internal non-compact (warped) Calabi-Yau are natural in the context of the gauge/string theory correspondence. We may consider for example the Klebanov-Strassler (KS) solution \cite{ks} and take its S-dual.\footnote{Since the dilaton is constant, we can tune it to be small in the S-dual solution.} The resulting solution is essentially given by a deformed conifold with $N$ units of $H$-flux on the $S^3$ deformation of the tip of the cone. The $S^3$ is a SLag cycle and thus a domain wall will consist of a D5-brane wrapping the $S^3$ and $N$ D3-branes ending on it.
Since the $H$-flux preserves the $SO(4)$ isometry group of $S^3$, as in the case of the $T^6/\mathbb{Z}_2$ orientifold, it can be considered to be constant on $S^3$. Thus it does not contribute to the right-hand side of (\ref{gensol}), since Eq.~(\ref{lapldw}) practically reduces to Eq.~(\ref{green}), where $H|_\Gamma$ gives the uniform compensating charge density $1/\hat V$. Alternately these can be seen directly by writing the usual expression of the Green's function for $S^3$ in terms of spherical harmonics. We do not try to make explicit this general expression for the Green's function, even if doable in principle, since it would be complicated and not particularly illuminating.    

We prefer to directly solve the BPS equation (\ref{bps}) in a symmetric case, already considered in \cite{kpv}, in which the D3-branes all end on the same point on the $S^3$.   Let us introduce the coordinates $(\psi,\phi,\chi)$ on $S^3$, such that the metric on $S^3$ takes the form 
\bea
\d s^2_{S^3}=R^2(\d\psi^2+\sin^2\psi\d \phi^2+\sin^2\psi\sin^2\phi\,\d\chi^2)
\eea
and  
\bea
H|_{S^3}=\frac{N}{2\pi^2}\sin^2\psi\sin\phi\, \d\psi\wedge\d\phi\wedge\d\chi\ .
\eea
The symmetry of the configuration allows only the ansatz
\bea
\calf=f(\psi)\sin\phi\,\d\phi\wedge\d\chi
\eea
for the worldvolume field strength. Let us place the $N$ D3-branes at the south pole $\psi=\pi$. As in \cite{kpv}, using Gauss' Law, the presence of the $N$ D3-brane magnetic sources at $\psi=\pi$ reduces to the boundary condition $f(\pi)=N/(4\pi)$, while $f(0)=0$.  The Bianchi identity $\d\calf=H|_{S^3}$ for $0\leq\psi<\pi$ reduces to $\partial_\psi f=\frac{N}{2\pi^2}\sin^2\psi$, which can be easily integrated
\bea
f(\psi)=\frac{N}{8\pi^2}(2\psi-\sin 2\psi)\ ,
\eea
satisfying the required boundary conditions. 

Since $X$ also depends only on $\psi$, the BPS equation (\ref{bps}) reduces to
\bea\label{dwequation}
\frac{\d X}{\d \psi}=\frac{f(\psi)}{R\sin^2\psi}=\frac{N}{8\pi^2R}\left(\frac{2\psi-\sin 2\psi}{\sin^2\psi}  \right)\ .
\eea
This is exactly the equation obtained in \cite{kpv} using an effective one-dimensional action for $\psi(X)$ where $X$ plays the role of the time, and using the conservation of the associated energy. We see how our general BPS equation (\ref{bps}) gives directly this result in this specific subcase. Equation (\ref{dwequation}) can be integrated
\bea
X=-\frac{N}{4\pi^2R}[\psi\,{\rm cot}\psi-1]+X_0\ ,
\eea 
where $X_0$ denotes the location of the domain wall, that is reached by $X$ at $\psi=0$.

The above domain walls  correspond in the dual gauge theory to the BPS domain walls interpolating between the baryonic and mesonic branches \cite{kpv}  and we have thus found the general formula which determines the shape of such domain walls for any choice of vacuum  in the moduli space of the mesonic branch.  Furthermore, as discussed in \cite{gkp},  the KS solution can be used to model  highly warped throat regions inside proper flux compactifications. Thus the above discussion can be adapted to these compactification scenarios as well, where the probe approximation may be better motivated than it was in the $T^6/\mathbb{Z}_2$ flux vacua considered in subsection \ref{Torient}.

\section{Conclusion and discussion}

There are two popular topological\footnote{The derived category classification captures more information than just the topology of the embedding, and it only applies to certain supersymmetric backgrounds.} D-brane classification schemes.  D-branes may be classified by homology, which captures some of the topological information about their embeddings, but misses all information about their gauge fields and in particular their lower-dimensional D-brane charges.  They may also be classified by twisted K-theory, which captures all of the RR charge information, both in the embedding and in the full nonabelian gauge field configuration, but with the disadvantage that given a particular D-brane configuration it is in general very difficult to determine the corresponding twisted K-theory class.  

In this note we have presented a classification scheme which lies between these two, the homology of generalized chains.  A generalized chain carries the information about the embedding of a D-brane and its worldvolume $U(1)$ gauge field, which allows one to treat D$p$-branes dissolved in and ending on D$(p+2)$-branes, and more generally networks of branes.  It has the advantage that it is described directly in terms of the physical data, the position and the fluxes, and so given a configuration one may immediately determine which homology class it represents, if any. Furthermore, the use of generalized chains is the most natural when background fluxes are turned on, since in the presence of a non-trivial $H$ the gauge-invariant worldvolume field strength $\calf$ can play a nontrivial role that must be considered along with with the embedding data. We introduced a boundary operator $\hat\partial$ for generalized chains and found that consistent D-brane networks wrap $\hat\partial$-closed generalized chains,  i.e.  generalized cycles. Therefore  they represent $\hat\partial$-homology classes.  We also saw that distinct representatives of the same homology class are related by allowed processes, sometimes involving the dissolution of branes in higher-dimensional ones, the nucleation of branes of various dimensions or the Myers effect.  Thus each homology class corresponds to a realizable value of the conserved RR charge, and vice versa.

Let us stress that, when considering stacks of D-branes, a full nonabelian description would in principle be required and  could lead to nontrivial effects. This is clearly not covered by our formalism and  we are in fact restricting to configurations where these effects  are neglected. However, let us note that several nonabelian effects have an alternate abelian description, which may be sufficient for some purposes.  For example, gauge bundles may have nontrivial instanton numbers.  While this is critical in the twisted K-theory classification of branes, in the $\hat\partial$-homology classification, we expect to lose no generality by considering only abelian bundles, as the instanton charges in the worldvolume of a D$p$-brane may also be carried by D$(p-4)$-branes outside of the D$p$-brane. On the other hand the inclusion of the abelian field strength, which couples to D$(p-2)$-brane charge, is critical because, as described in \cite{mms} and reviewed in sec.~\ref{genhom}, Dirac monopoles in time-dependent D$p$-brane configurations violate D$(p-2)$-brane charge conservation.

While a physical process may transform any representative of a fixed homology class into any other, such a process may not be energetically favorable, and so will not occur at zero temperature.  In the second half of this note we describe how the calibrations introduced in \cite{luca1,paulk} can be extended and are in fact more naturally associated to  D-brane networks.   These calibrations are polyforms that provide a local bound on the energy density of a D-brane network, a bound that is everywhere saturated precisely for supersymmetric configurations. Integrated over a generalized cycle, the calibration provides the supersymmetric lower bound for the energy of the D-brane network in the associated $\hat\partial$-homology class.  

In sections~\ref{fluxsec} and \ref{cysec}  we have focused on the case of flux compactifications to four-dimensions and we constructed explicit examples of networks of domain walls, strings and space-filling branes in the toroidal orientifold backgrounds of \cite{kachru}.  In section~\ref{2exsec} we explicitly found the geometry of domain wall configurations in IIB warped Calabi-Yau compactifications, where a number of space-time filling D3-branes end on a D5-brane wrapping an internal three-cycle which supports a non-trivial $H$-flux. In particular, by using the appropriate calibration (\ref{wcycal}) we obtained the explicit BPS equations (\ref{DWcond}) and (\ref{bps}) describing the resulting composite domain walls. The same procedure can be applied to slightly different settings and of course one can switch off the fluxes and consider pure Calabi-Yau backgrounds.  For example,  our formalism may be used to describe the setting of \cite{verlinde}, in which a  D5-brane wraps an almost supersymmetric two-cycle at the bottom of a conifold-like geometry that can unwrap by sweeping out a three-cycle of large volume inside the whole compact Calabi-Yau.  From the appropriate generalized calibration one can obtain BPS equations (analogous to (\ref{DWcond}) and (\ref{bps})) for the interpolating domain-wall, that can be  useful in the description of these kinds of decays.

In this note we have stressed the usefulness of the homology of generalized chains on the spacetime.  However one is often interested in the spectra of various branes in the presence of certain fixed background branes.  For example, consider the Hanany-Witten cartoons whose D4 and NS5-branes are described by MQCD.  The excitations of MQCD correspond to D2-branes, F-strings, D0-branes and networks made out of all three, describing for example magnetic monopoles confined by vortices \cite{necklace,robertostefano,ST}.  In these cases noncomposite objects are classified by the relative homology of the spacetime with respect to the fixed D4 and NS5-branes \cite{WittenMQCD2,Stefano}, while the composite objects will be classified by the relative homology of generalized chains.  

In summary, the following holes need to be filled.  A relative homology theory should be defined, we expect this to be straightforward.  We have not included gravitational effects, which naively corresponds to adding an $A$-roof genus in the definition of the generalized currents.  Our formalism does not treat nonabelian worldvolume gauge fields, these should change the energy formula which determines the calibrations, as $\calf$ and the embedding fields become matrix-valued.  Our formalism is not S-duality covariant, S-duality covariance requires information about the dual gauge bundle.  Finally, our treatment is largely classical as it uses differential forms, the Dirac quantization condition is included by hand and so torsion classes and anomalies are missing.

\vspace{1cm}

\section*{Acknowledgements}
We thank R.~Argurio, F.~Denef, P.~Grange, P.~Koerber, V.~Mathai, W.~Troost and R.~Zucchini for useful discussions.  L.~M.\ is supported in part by the Federal Office for Scientific, Technical and Cultural Affairs through the ``Interuniversity Attraction Poles Programme -- Belgian Science Policy" P5/27 and by the European Community's Human Potential Programme under contract
MRTN-CT-2004-005104 `Constituents, fundamental forces and symmetries of the universe'. 

\vspace{1cm}

\begin{appendix}

\section{Topology and quantum effects}
\label{torsion}

In this appendix we note that the twisted homology theory of \cite{Andres} captures quantum corrections that are not present in this note.  We argue that while that theory does not apply in certain cases, a correction may be inserted by hand that fixes it in general.  We postulate the existence of a purely geometric formation of $\hat\partial$-homology which captures the quantum effects of twisted homology together with this quantum correction.

We have introduced a $\hat\partial$-homology for generalized cycles which is not manifestly integral. We expect that integrality would arise automatically in a formalism in which the worldvolume gauge field is quantized.  However if the gauge field is treated as a differential form, then one will never capture configurations in which branes with torsion ($\Z_n$) charges are dissolved in higher-dimensional branes.   Such configurations are included in manifestly integral twisted homology theories like that of \cite{Andres}, which we expect is rationally isomorphic to the $\hat\partial$-homology presented here.

Integral twisted homology is defined on ordinary chains using the $H$-twisted boundary operator
\bea \label{idiff}
\partial_H x=\partial x+H\cap x\ .
\eea
Here the $H$ flux is an integral 3-cocycle, and the cap product $\cap$ is an operation which takes a $p$-cocycle and a $q$-chain and yields a $(q-p)$-chain, which corresponds to the worldvolume of our magnetic monopole.   We refer to \cite{Andres} for further details on the definition of twisted homology, but in this appendix we will extend this formalism to include configurations with a nontrivial Freed-Witten anomaly, which were explicitly excluded in \cite{Andres} because $\partial_H$ ceases to be nilpotent in these cases.

While $\partial_H$ is not nilpotent in general, the product $H\cup H$ is always nilpotent
\bea
2H\cup H=0
\eea
and so it vanishes in de Rham cohomology, which contains no torsion elements.  In terms of differential forms this is reflected in the fact that
\bea
H\wedge H=0
\eea
as $H$ is an odd-dimensional form. 

This is not to say that string theory is inconsistent at the level of integral homology when $H\cup H$ is nontrivial.  In the full quantum theory a magnetic monopole worldvolume $\calc$ is not only sourced by the pullback of the $H$ flux to the worldvolume $\Sigma$, but one must also add a quantum correction equal to the third Stiefel-Whitney class $W_3$ of the tangent bundle of $X$ \cite{FW}.  The magnetic monopole charge on $\calc$ is then not only the pullback of $H$ from $\Sigma$, which is nonzero when $H\cup H\neq 0$ in $\Sigma$.  It also contains a contribution $W_3$ of the normal bundle to $\calc$ in $\Sigma$.  This always precisely cancels the worldvolume $H$ flux on $\calc$.  Therefore the brane wrapped on $\calc$, which must be inserted to restore gauge invariance to a brane wrapping $\Sigma$, is itself not anomalous, because its total Freed-Witten anomaly is
\bea
W_3+H=0.
\eea
Thus while $\partial+H\cup$ is not nilpotent when $H\cup H\neq 0$, if one adds the quantum correction $W_3$ it because nilpotent and the homology theory is well-defined.  

$W_3$ of the normal bundle of $\calc$ in $\Sigma$ is only defined on a cycle, and $\calc$ may not be a cycle.  However its pushforward to $\Sigma$ is equal to $Sq^3 H$, where $Sq^3$ is an operation known as a Steenrod square, as it squares 3-cocycles.  If we define $Sq^3$ on a 3-cochain $H$ to be the cup product $H\cup H$, then it is defined even for cochains and in the case of cocycles it reproduces the Freed-Witten term $W_3$.  Now the monopole worldvolume on $\calc$ is no longer just
\bea
H\cap\calc=H\cap(H\cap\Sigma)=(H\cup H)\cap\Sigma
\eea
but it contains $Sq^3$, acting as described above, and so is
\bea
&&(Sq^3+H\cap)\calc=(Sq^3+H\cap)(H\cap\Sigma)=\cr &&=(Sq^3 H+H\cup H)\cap\Sigma=(2H\cup H)\cap\Sigma=0\cap\Sigma=0.
\eea
Using this prescription one finds that twisted homology is equal to the second element in the Atiyah-Hirzebruch spectral sequence construction of K-theory, of which $Sq^3+H$ is the first differential.  In the present note we do not need this quantum correction because we consider differential forms, and as a differential form $H\wedge H=0$ always.  

It would be interesting to analogously extend the $\hat\partial$-homology of this note to a purely topological construction, working directly with the worldvolume twisted gauge bundles and $spin^c$ bundle.  In such a formulation one may hope to naturally obtain a complete quantum corrected boundary operator, thus including the $W_3$ quantum correction introduced above by hand.


\section{Flux quantization on generalized cycles}
\label{RRquant}

The formalism introduced in this paper allows one to naturally express the quantization condition on the Ramond-Ramond fluxes.  Indeed, we have seen in subsection \ref{omrit} that a consistent D-brane network must wrap a generalized cycle $(\mathfrak{C},\mathfrak{F})$ (such that $\hat\partial(\mathfrak{C},\mathfrak{F})=0$). Using the current $j_{(\mathfrak{C},\mathfrak{F})}$, the CS-term in the network's action is given by
\bea
S_{\rm CS}=\int_X\langle C, j_{(\mathfrak{C},\mathfrak{F})}\rangle\ .
\eea

However, since the RR gauge field polyform $C$ is defined only locally, it is convenient to express $S_{\rm CS}$  in terms of the globally defined field strength $F=\d_H C$. This can be done by choosing a fixed generalized cycle $(\mathfrak{C}_0,\mathfrak{F}_0)$ in the same generalized homology class as $(\mathfrak{C},\mathfrak{F})$, so that there exists a generalized chain $(\fraks,\hat\frakf)$ such that $\hat\partial(\fraks,\hat\frakf)=(\mathfrak{C},\mathfrak{F})-(\mathfrak{C}_0,\mathfrak{F}_0)$. Then we can write
\bea
S_{\rm CS}(\mathfrak{C},\mathfrak{F})=S_{\rm CS}(\mathfrak{C}_0,\mathfrak{F}_0)+\int_X\langle F, j_{(\fraks,\hat\frakf)}\rangle\ .
\eea
In our units  $2\pi\sqrt{\alpha^\prime}=1$ and so this term enters the path integral via the exponential
\bea
\exp(2\pi iS_{\rm CS})\ ,
\eea
which must be independent of the choice of $(\fraks,\hat\frakf)$. Clearly, any two such generalized chains $(\fraks,\hat\frakf)$ and $(\fraks^\prime,\hat\frakf^\prime)$ differ by a generalized cycle   $(\tilde{\mathfrak{C}},\tilde{\mathfrak{F}})$ and so we obtain the quantization condition 
\bea
\int_X\langle F, j_{(\tilde{\mathfrak{C}},\tilde{\mathfrak{F})}}\rangle\in \mathbb{Z}
\eea
for any generalized cycle   $(\tilde{\mathfrak{C}},\tilde{\mathfrak{F}})$. The above quantization condition depends only on the $\d_H$-cohomology class of $F$ and the $\hat\partial$-homology class of $(\tilde{\mathfrak{C}},\tilde{\mathfrak{F}})$, and therefore it is a properly defined topological property. 

We have thus obtained a characterization of the RR-flux quantization condition in terms of generalized cycles, which allows one to naturally generalize  standard arguments to the most general case where in particular one can have non-trivial $H$-flux. Note that the above argument is quite naive and does not explicitly face possible subtleties due to anomalous corrections, torsion effects, orientifolds, etc. It would be nice to address these and other related issues in the future.   

\end{appendix}


\end{document}